\documentclass[11pt]{article}
\usepackage{fancyhdr}
\usepackage{isomath}
\usepackage{amsmath}
\usepackage{amsbsy}
\usepackage{amssymb}
\usepackage{amscd}
\usepackage{amsfonts}
\usepackage{graphicx,color}
\usepackage{verbatim}
\usepackage{euscript}
\usepackage{alltt}
\usepackage{stmaryrd}
\usepackage{subfigure}
\usepackage{url}

\newtheorem{definition}{Definition}
\newtheorem{assumption}{Assumption}

\usepackage{amsmath}
\usepackage{amsbsy}
\usepackage{amssymb}
\usepackage{amscd}
\usepackage{amsfonts}

\newcommand{\R}{\mathbb R}

\newcommand{\bfa}{{\mathbold a}}
\newcommand{\bfb}{{\mathbold b}}
\newcommand{\bfc}{{\mathbold c}}

\newcommand{\bfe}{{\mathbold e}}

\newcommand{\bfl}{{\mathbold l}}

\newcommand{\bfn}{{\mathbold n}}

\newcommand{\bfp}{{\mathbold p}}
\newcommand{\bfq}{{\mathbold q}}
\newcommand{\bfr}{{\mathbold r}}
\newcommand{\bfs}{{\mathbold s}}
\newcommand{\bft}{{\mathbold t}}
\newcommand{\bfu}{{\mathbold u}}

\newcommand{\bfx}{{\mathbold x}}
\newcommand{\bfy}{{\mathbold y}}
\newcommand{\bfz}{{\mathbold z}}

\newcommand{\bfA}{{\mathbold A}}

\newcommand{\bfC}{{\mathbold C}}

\newcommand{\bfF}{{\mathbold F}}

\newcommand{\bfH}{{\mathbold H}}
\newcommand{\bfI}{{\mathbold I}}

\newcommand{\bfQ}{{\mathbold Q}}
\newcommand{\bfR}{{\mathbold R}}

\newcommand{\bfU}{{\mathbold U}}

\newcommand{\bfW}{{\mathbold W}}

\newcommand{\veps}{\varepsilon}

\newcommand{\beq}{\begin{equation}}
\newcommand{\eeq}{\end{equation}}
\newcommand{\beqs}{\begin{eqnarray}}
\newcommand{\eeqs}{\end{eqnarray}}
\newcommand{\beql}{\begin{equation} \label}
\newcommand{\half}{\frac{1}{2}}

\newtheorem{theorem}{Theorem}[section]
\newtheorem{lemma}{Lemma}[section]

\newtheorem{remark}{Remark}[section]

\newcommand{\bfomega}{\mathbold{\omega}}

\newcommand{\bfOmega}{\mathbold{\Omega}}

\newcommand{\bfveps}{\mathbold{\varepsilon}}

\newcommand{\bfzero}{\mathbf{0}}

\newcommand{\lsup}[2]{{^ #1 #2}}

\usepackage[margin=1in]{geometry}

\date{October 14, 2017}
\begin{document}
\title{On Weingarten-Volterra defects{\footnote{\textbf{To appear in Journal of Elasticity.}}}}
\author{Amit Acharya\thanks{Dept. of Civil \& Environmental Engineering, and Center for Nonlinear Analysis, Carnegie Mellon University, Pittsburgh, PA 15213, email: acharyaamit@cmu.edu.} 
}
\maketitle

\begin{abstract}
\noindent The kinematic theory of Weingarten-Volterra line defects is revisited, both at small and finite deformations. Existing results are clarified and corrected as needed, and new results are obtained. The primary focus is to understand the relationship between the disclination strength and Burgers vector of deformations containing a Weingarten-Volterra defect corresponding to different cut-surfaces.
\end{abstract}

\section{Introduction}
The question of characterizing the discontinuity of a deformation whose strain is locally compatible with a prescribed field on simple types of non-simply-connected domains is the main concern of this paper. Such questions originated in the works of Weingarten \cite[as translated by \cite{delph_wein}]{wein} and Volterra \cite[as translated by \cite{delph_volt}]{volt} in the setting of small deformations, and in those of Zubov \cite{zubov} and Casey \cite{casey} in the context of finite deformations; related work is that of Yavari \cite{yav}, considering compatibility  conditions, i.e., conditions for continuous deformations for a prescribed strain field, in general non-simply connected domains. Both  finite and small deformations (i.e., linearized kinematics) are considered. A motivation for this paper is the recent emphasis on developing and understanding models of defects in materials.

We revisit classical results, provide alternative proofs, and correct some statements in the existing literature related to uniqueness of the disclination strength and Burgers vector of defects. We also provide new results related to the dependence of these objects on cut-surfaces. 

A natural improvement of the work presented herein is to deduce the corresponding results for arbitrary non-simply connected domains, and make precise connections with the results of the metric differential geometric treatment of Kupferman, Moshe, and Solomon \cite{kupferman2015metric}. Such a connection is desirable, as the differential geometric treatment does not involve notions of deformations of bodies and their discontinuities, while the continuum mechanics point of view, starting from Weingarten \cite{wein}, Volterra \cite{volt}, and Zubov \cite{zubov}, is deeply rooted in the kinematics of deformation of 3-d bodies.

After this brief introduction, Section \ref{sec:setting} provides the setting of the main questions asked in the paper. Section \ref{sec:small} considers the questions in the setting of small deformations. Section \ref{sec:large} considers the same questions for kinematics without approximation. The paper contains an Appendix collecting classical results on compatibility on simply-connected domains.

A somewhat mathematical style of presentation is adopted simply for the purpose of a better organization of definitions, assumptions, results, and remarks.

\section{The setting and the question of Weingarten's theorem}\label{sec:setting}
\begin{definition}\label{def:region}
By a region $\Omega$ we will mean a pathwise-connected open set in ambient 3-d Euclidean point space ${\cal{E}}_3$, together with some or all of its boundary points  \cite{kellog}. In contrast to standard continuum mechanics, we will need to consider non-compact bounded regions, generated by removing surfaces from compact regions.
\end{definition}
\begin{definition}\label{def:defmn}
By a deformation $\bfy:\Omega \rightarrow V_3$ we mean a $C^1$ mapping $\bfy$ of $\Omega$ to the translation space $V_3$ of ${\cal{E}}_3$, with pointwise positive determinant of its gradient, i.e., $\det(grad\,\bfy) > 0$. The displacement is defined as $\bfu(\bfx) := \bfy(\bfx) - \bfx$ for all $\bfx \in \Omega$.
\end{definition}
Given a \emph{simply connected} region $\Omega$ and a prescribed  twice continuously-differentiable, positive-definite, symmetric tensor field $\bfC$  (a symmetric second order tensor field $\bfveps$) on it, it is a classical result of continuum mechanics, e.g., \cite{shield}, that a thrice-differentiable deformation (displacement) field can be constructed on it whose Right Cauchy-Green deformation (strain) tensor is the prescribed field $\bfC$ ($\bfveps$), if the Riemann-Christoffel curvature tensor formed from $\bfC$ (the St.-Venant tensor formed from $\bfveps$) vanishes, i.e., 
\begin{equation}\label{eqn:loc_comp}
\begin{split}
& \Gamma^\gamma_{\alpha \beta} := \frac{1}{2} (C^{-1})^{\gamma \mu}\left[ C_{\alpha \mu, \beta} +C_{\beta \mu, \alpha} - C_{\alpha \beta, \mu}\right],\\ & R^\mu_{\alpha \beta \rho} := \Gamma^\mu_{\alpha \beta, \rho} - \Gamma^\mu_{\alpha \rho, \beta} +\Gamma^\mu_{\gamma \rho} \Gamma^\gamma_{\alpha \beta} + \Gamma^\mu_{\gamma \beta} \Gamma^\gamma_{\alpha \rho} = 0,\\
& \veps_{il,km} - \veps_{kl,im} - \veps_{im,kl}+ \veps_{km,il} = 0.
\end{split}
\end{equation}

The main purpose of Weingarten's theorem may be stated as understanding the obstruction to the above-mentioned construction of the deformation (displacement) field when the region is no longer simply-connected. 
\begin{definition} 
When $\bfC$ ($\bfveps$) satisfies conditions \eqref{eqn:loc_comp}, we refer to it as locally compatible.
\end{definition}
\begin{definition}
Given a $\bfC$ ($\bfveps$) field on a region, any deformation $\bfy$ (displacement $\bfu$) of the region that satisfies $(grad\, {\bfy})^T grad\, {\bfy} = \bfC$ $((grad\, {\bfu})_{sym} = \bfveps)$ is said to be (strain) compatible with $\bfC$ ($\bfveps$) on the region.
\end{definition}
\begin{definition}\label{def:rigid_large}
Two deformations $\bfy_1$ and $\bfy_2$ of a region $\Omega$ are related by a rigid deformation if there exists a (proper) orthogonal tensor $\bfR$, constant on $\Omega$, such that $\bfy_1(\bfx) - \bfy_1(\bfz) = \bfR\left[ \bfy_2(\bfx) - \bfy_2(\bfz) \right]$ for all $\bfx, \bfz \in \Omega$. 
\end{definition}
We note that assuming $\bfR$ to be a `small' rotation in Definition \ref{def:rigid_large} so that $\bfR \approx \bfI + \bfW$, with $\bfW$ skew,  results in the statement $\bfu_1(\bfx) - \bfu_1(\bfz) = \bfu_2(\bfx) - \bfu_2(\bfz) + \bfW \left[ \bfx - \bfz \right] + \bfW \left[ \bfu_2(\bfx) - \bfu_2(\bfz) \right]$ for all $\bfx, \bfz \in \Omega$. In what follows, we will make the further assumption  that $\left|\bfW \left[ \bfu_2(\bfx) - \bfu_2(\bfz) \right] \right|$ is small  and define

\begin{definition}\label{def:rigid_small}
Two displacements $\bfu_1$ and $\bfu_2$ of a region $\Omega$ are related by an infinitesimally rigid deformation if there exists a skew symmetric tensor $\bfW$, constant on $\Omega$, such that $\bfu_1(\bfx) - \bfu_1(\bfz) = \bfu_2(\bfx) - \bfu_2(\bfz) + \bfW \left[ \bfx - \bfz \right]$ for all $\bfx, \bfz \in \Omega$. 
\end{definition}
\begin{remark}\label{rem:rigid_const}
Given two deformations related to each other by a (infinitesimal) rigid deformation, it is often analytically convenient to view the rigidity statement in the form $\bfy_1(\bfx) = \bfR \bfy_2(\bfx) + \bft$  $\left( \bfu_1(\bfx) = \bfu_2(\bfx) + \bfW\bfx + \tilde{\bft}\, \right)$, where $\bft := \bfy_1(\bfz) - \bfR \bfy_2(\bfz)$  $(\tilde{\bft} := \bfu_1(\bfz) - \bfu_2(\bfz) - \bfW \bfz )$ for any $\bfz \in \Omega$ is a constant vector on $\Omega$. However, it should be kept in mind that the constant $\bft$ $(\tilde{\bft})$ so defined is not independent of the choice of the origin chosen to define position vectors when $\bfR \neq \bfI$ $(\bfW \neq \bfzero)$. Therefore it is not a constant in the physical sense, while the fundamental definition of a (infinitesimally) rigid deformation is a physical statement independent of the choice of an origin. This seemingly trivial point is of some importance in this paper (see Remark \ref{rem:burgers_small_1} and Sec. \ref{subsec:burgers_large}).
\end{remark}

\begin{figure}
\centering
\includegraphics[width=6.5in, height=3.0in]{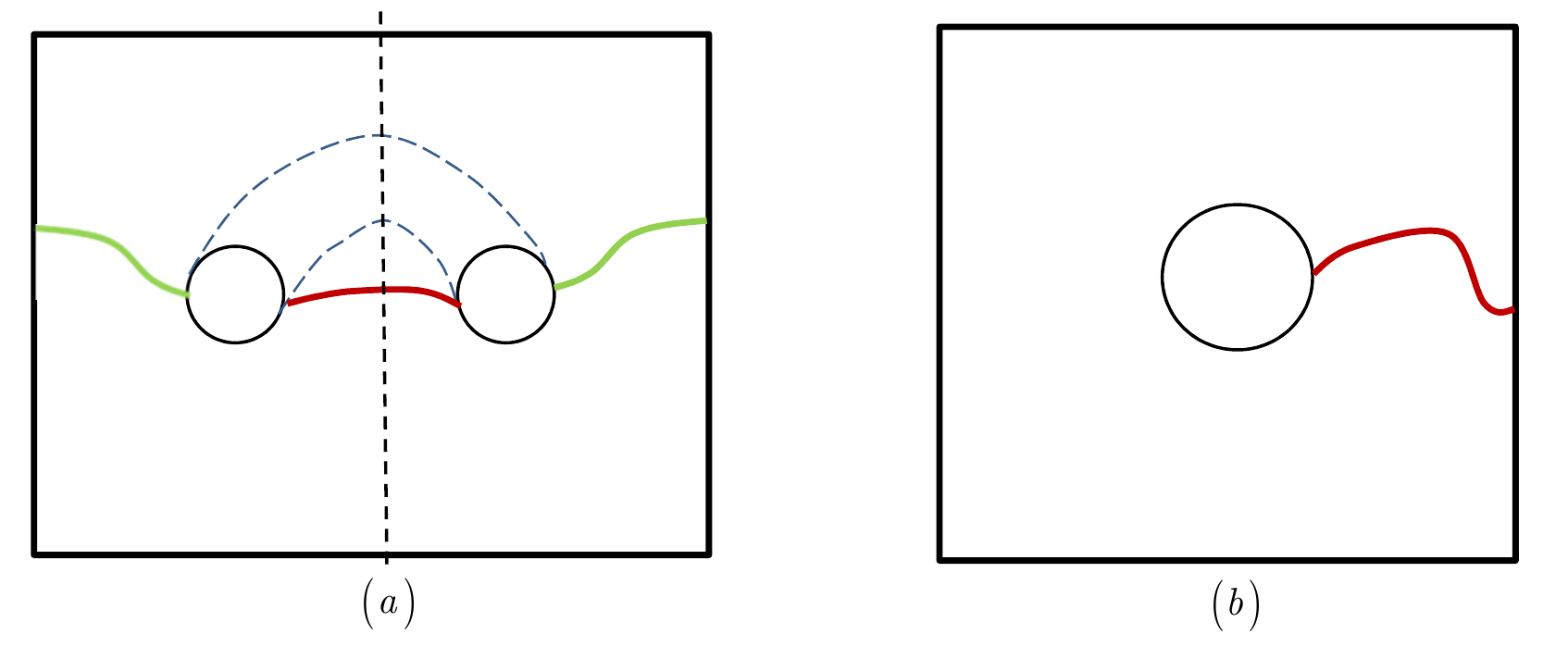}
\caption{Cross-section of a body with  a (a) toroidal and (b) through hole. (a) The whole body may be thought of as generated from a rotation of the cross-section (with no colored lines) by $\pi$ radians about the vertical dashed line. The curved dashed lines are intended to provide a rough perspective idea of the toroidal hole behind the cross-section. The red curve is the trace of one possible cut-surface and the green one, that of another; each cut renders the body simply connected. The red cut surface produces a topological ball with a cavity. The green cut produces a topological ball. To imagine some simply connected bodies corresponding to either the red cut-surface or the green one, think of the cross-section with the corresponding colored trace being rotated by $\pi$ radians about the vertical dashed line. (b) The whole body is generated by extruding the cross-section perpendicular to itself. The red curve is the trace of a surface generated by extruding the curve. Extracting the red cut surface from the cylinder with a through hole produces a simply connected topological ball.}
\label{fig:fig_bodies}
\end{figure}

It is a classical result, see, e.g., \cite{shield}, that if two continuous deformations (displacements) on the same region have identical, continuous right Cauchy-Green (strain) fields, then one is at most a rigid (infinitesimally rigid) deformation of the other. For the sake of completeness, we provide the main elements of the proofs  of these classical results in Appendix \ref{comp_sc}.
\begin{definition}
By a (cut)-surface of $\Omega$ we mean a 2-d set of points in $\Omega$ lending itself to a smooth parametrization from a collection of (often one) squares of $\R^2$ (that can be smoothly mapped to each other with orientation preserved), that provides a natural sense of orientation of the surface (through the choice of normal constructed from the parametrization). All surfaces will be assumed to be non self-intersecting.
\end{definition}
In what follows, we will consider two elementary types of non-simply connected regions. One will be a 3-dimensional body with a through-hole such that there are curves in the body that cannot be continuously shrunk to a point while staying within the body. Removing a \emph{cut-surface} from the body connecting the inner hole to the outer boundary can render the body simply-connected, with the topology of a ball. The other type of non simply connected body is one with a toroidal hole in it. Removing a cut-surface from the exterior boundary of the body to the boundary of the hole again renders the body simply connected, the resulting body having the topology of a ball. Another alternative is to remove a cut-surface in the body that changes the toroidal hole into an opening with the topology of a connected cavity. See Fig. \ref{fig:fig_bodies} for illustration of these concepts, also see \cite[p.16]{nabarro}.  
\begin{assumption}\label{def:region}
We consider any non-simply connected region $\Omega$ that is reduced to a simply connected region $\Omega_\tau := \Omega \backslash \tau$ by the removal of a single cut-surface, $\tau$.
\end{assumption}
\begin{assumption}
For $\bfx \in \tau$ and a function ${^\tau f}$ defined on $\Omega_\tau$, we will assume that unique limits $\lim_{\bfx^\pm \rightarrow \bfx} {^\tau f} (\bfx^\pm)$,  with $\bfx \in \tau$ and $\bfx^\pm$ approaching $\bfx$ from either side of the surface $\tau$, exist; we will denote these limiting values as ${^\tau f}^\pm(\bfx)$ for each $\bfx \in \tau$. We will also use the notation
\[
\llbracket{^\tau f} (\bfx) \rrbracket := {^\tau f}^+(\bfx) - {^\tau f}^-(\bfx).
\]
We think of a sequence of points approaching $\bfx \in \tau$ from a `side' in the intuitively natural way. If $\bfn(\bfx)$ is the unit normal to $\tau$ at $\bfx$ (arbitrarily choosing one alternative), we think of the sequence $\{\bfx^\pm_i\}$ as approaching $\bfx$ from the $\pm$ side if $(\bfx^\pm_i - \bfx)\cdot \bfn(\bfx) \gtrless 0$ for all $i$.
\end{assumption}

The main question addressed by {\bf\emph{Weingarten's}} theorem may now be stated as follows:
\begin{quote}
Given a non-simply connected region $\Omega$ as described above and a twice continuously differentiable,  positive-definite, symmetric, locally compatible tensor field $\bfC$  (a symmetric second order tensor field $\bfveps$) on it, characterize the `jump' $\llbracket {^{\tau}\bfy}(\bfx) \rrbracket$ $\left(\llbracket ^\tau\bfu (\bfx) \rrbracket \right)$, of any deformation ${^\tau \bfy}$ (displacement ${^\tau \bfu}$) field compatible with $\bfC$ $(\bfveps)$ that can be constructed on $\Omega_\tau$. In particular, we will be interested in understanding to what extent the characterization of this jump is independent of points $\bfx$ on a fixed cut-surface $\tau$ and to what extent the jump functions across different cut-surfaces may be related.
\end{quote}
\begin{definition}
Given a cut-surface $\tau$ of $\Omega$ and a $\bfC$ $(\bfveps)$ field on $\Omega$, we refer to the latter's restriction to $\Omega_\tau$ as $\bfC_\tau$ $(\bfveps_\tau)$.
\end{definition}

\begin{remark}
We note that the construction of a family of thrice continuously differentiable deformations (displacements) with $(grad\, {^\tau\bfy})^T grad\, {^\tau\bfy} = \bfC_\tau$ $((grad\, {^\tau \bfu})_{sym} = \bfveps_\tau)$ on any $\Omega_\tau$ is guaranteed; however, because of the non-simply connectedness of $\Omega$, the limits of such a deformation (displacement) at points of the cut-surface $\tau$ from either side of $\tau$ may not match. The goal of the Weingarten theorem is to characterize the discontinuity of the deformation (displacement), when viewed as a function on $\Omega$. 
\end{remark}
\begin{definition}
 We say that a deformation ${^\tau \bfy}$ (displacement ${^\tau \bfu}$) on $\Omega_\tau$ contains a \textbf{Weingarten-Volterra defect} if it displays a non-vanishing jump $\llbracket {^\tau \bfy} \rrbracket$ $(\llbracket {^\tau \bfu} \rrbracket)$.
\end{definition}
\begin{remark}\label{rem:casey}
The role of any cut-surface $\tau$ in our considerations is to produce a simply-connected region from $\Omega$. If $\Omega$ was simply connected to begin with, one could consider removing a cut-surface $\tau$ from it, but only of the type that would keep $\Omega_\tau$ simply-connected (this has physical importance in keeping the dislocation line within the body). It is clear from classical arguments (Appendix \ref{rigid_large}) that any two differentiable deformations of $\Omega_\tau$ with identical Right Cauchy Green fields are related to each other by at most a rigid deformation. Given a locally compatible $\bfC$ field on $\Omega$, any deformation, say ${^\tau \bfy}$, compatible with $\bfC_\tau$, would necessarily differ from the restriction of  any deformation $\bfy$ compatible with $\bfC$ on $\Omega$ to $\Omega_\tau$ only by a rigid deformation of ${\bfy}(\Omega_\tau)$, i.e., ${^\tau\bfy}(\bfx^\pm) = \bfR \bfy(\bfx^\pm) + \bft$ for $\bfx^\pm \in \Omega_\tau$, for some orthogonal tensor $\bfR$ and vector $\bft$, both constant on $\Omega_\tau$. Passing to the limit $\bfx^\pm \rightarrow \bfx \in \tau$ with sequences $\bfx^+$ and $\bfx^-$ approaching $\bfx$ from either side of the cut, we have that $\llbracket {^\tau \bfy} \rrbracket = \bfzero$.
Thus, it is impossible for a deformation of a simply connected region $\Omega_\tau$ induced from a simply connected $\Omega$ to display a Weingarten-Volterra defect if it is compatible, on $\Omega_\tau$, with a $\bfC_\tau$ induced from a locally compatible $\bfC$ field on $\bfOmega$. The `counterexample' that Casey provides \cite[Example 1, p. 485]{casey} for this result appears to be related to the fact that $\Omega_\tau$ in his example (induced from a simply connected $\Omega$) is not a path-connected region (allowed by his hypothesis adapted from \cite[p.42]{gurtin}), and therefore it would not be possible to conclude that $\bfy$ and ${^\tau\bfy}$ in our construction are necessarily related by a single rigid deformation for such situations (see Appendix \ref{rigid_large}).
\end{remark}
\begin{remark}
The essential content of the argument in Remark \ref{rem:casey} was important to Volterra (\cite[as translated in \cite{delph_volt}]{volt}; Volterra worked with the small deformation theory) in making the case that one could have a discontinuous elastic deformation only if the body was not simply connected or if its strain field contained a singularity in a simply connected domain (or both). We note that the motivation for the cut-surface in Volterra's arguments was to improve the topological situation by making a non-simply connected region into a simply-connected one (and not worse by taking a simply connected domain and making it disconnected by a through-cut-surface).
\end{remark}

\section{Small deformation}\label{sec:small}

\subsection{Weingarten's theorem for small deformation}\label{sec:wein_small}
We give a proof of Weingarten's theorem that involves different arguments from those presented in Love and Nabarro \cite{love, nabarro}.

Given a simply connected $\Omega_\tau$ (induced from the non-simply connected $\Omega$ with a cylindrical/toroidal hole) and a locally compatible $\bfveps$ field on it, consider any displacement field ${^\tau\bfu}$ compatible with $\bfveps$ on $\Omega_\tau$.  Appendix \ref{comp_small} shows that that there exists a family of such displacement fields, each member of which satisfies
\begin{equation}\label{eqn:rotg_straing}
\begin{split}
&\frac{1}{2} \left( {^\tau u}_{i,j} + {^\tau u}_{j,i} \right) = \veps_{ij},\\
&  \frac{1}{2} \left( {^\tau u}_{i,j} - {^\tau u}_{j,i} \right) =: {^\tau \omega}_{ij},\\
&{^\tau \omega}_{ik,l} = \veps_{il,k} - \veps_{kl,i} =: E_{ikl}
\end{split}
\end{equation}
on $\Omega_\tau$. Strictly speaking, the $\bfveps$ appearing in \eqref{eqn:rotg_straing} is $\bfveps_\tau$.
\begin{remark}\label{rem:jump_family}
Any two such displacement fields on $\Omega_\tau$ compatible with the same strain field necessarily differ by an infinitesimally rigid deformation and, therefore, it follows from Remark \ref{rem:rigid_const} that their jump fields across $\tau$ are necessarily equal. It also follows that the jump in their infinitesimal rotation field across $\tau$ is equal.
\end{remark}
\begin{definition}\label{def:curve}
For the purpose of this paper, we think of a curve as a 1-d set of points in $\Omega$ lending itself to a smooth parametrization from some interval in $\R$ which provides a natural sense of direction on the curve. All curves will be assumed to be non self-intersecting (i.e., simple curves).
\end{definition}

Consider a curve $\bfc_{(\bfx,\bfz)}$ on the surface $\tau$ joining $\bfx$ and $\bfz$. Corresponding to $\bfc$, consider two other curves $\bfc^+_{(\bfx^+,\bfz^+)}$ and $\bfc^-_{(\bfx^-,\bfz^-)}$ in $\Omega_\tau$ on either side of, and close to, $\tau$. The curves run from $\bfx^\pm$ to $\bfz^\pm$ (in obvious notation), and in the following we will be thinking of limits of line integrals along $\bfc^\pm$ as $\bfc^\pm$ tend to $\bfc$.

We may write
\begin{equation*}
\lsup{\tau u_i}(\bfz^\pm) - \lsup{\tau u_i}(\bfx^\pm) = \int_{\bfc^\pm_{(\bfx^\pm,\bfz^\pm)}} \left( \veps_{ij}(\bfx'^\pm) +   \lsup{\tau \omega_{ij}}(\bfx'^\pm) \right) \, dx'^\pm_j,
\end{equation*}
and taking the limit as ${\bfc^\pm_{(\bfx^\pm,\bfz^\pm)}} \rightarrow \bfc_{(\bfx,\bfz)}$ and then subtracting the $\pm$ equations we obtain
\begin{equation}\label{eqn:pre_wein}
\llbracket \lsup{\tau u_i} (\bfz) \rrbracket = \llbracket \lsup{\tau u_i} (\bfx) \rrbracket + \int_{\bfc_{(\bfx,\bfz)}} \llbracket \lsup{\tau \omega_{ij}} (\bfx') \rrbracket \, dx'_j,
\end{equation}
noting that the field $\bfveps$ is continuous at each point of $\bfc_{(\bfx, \bfz)}$ by hypothesis. Now, for each $\bfx' \in \tau$,  $\llbracket \lsup{\tau \omega_{ij}} (\bfx') \rrbracket$ can be expressed,using \eqref{eqn:rotg_straing}, as
\begin{equation}\label{eqn:rot_jump_ind}
\begin{split}
\llbracket \lsup{\tau \omega_{ij}} (\bfx') \rrbracket &= \lim_{\bfx'^\pm \rightarrow \bfx'} \int_{\bfc_{(\bfx'^-,\bfx'^+)}} ( \lsup{\tau \omega_{ij,m}} (\bfs) ) \,ds_m \\
& = \lim_{\bfx'^\pm \rightarrow \bfx'} \int_{\bfc_{(\bfx'^-,\bfx'^+)}} \left[ \veps_{im,j}(\bfs) - \veps_{jm,i} (\bfs) \right] \,ds_m = \int_{\bfl_{\bfx'}} \left[ \veps_{im,j}(\bfs) - \veps_{jm,i} (\bfs) \right] \,ds_m
\end{split}
\end{equation}
where $\bfc_{(\bfx'^-,\bfx'^+)}$ is a curve from $\bfx'^-$ to $\bfx'^+$ contained in $\Omega_\tau$, and $\bfl_{\bfx'}$ is a closed loop in $\Omega$ passing through $\bfx'$ that pierces $\tau$ exactly once (i.e., the loop $l_{\bfx'}$ goes around the hole in $\Omega$).
\begin{definition}
For our purposes, a non-contractible loop through $\bfx$ in a non-simply connected region $\Omega$ (as in Definition \ref{def:region}) is a closed curve in $\Omega$  that cannot be contracted to a point without exiting $\Omega$; and, it intersects only once some cut-surface containing $\bfx$.
\end{definition}
Thus, the loop $\bfl_{\bfx'}$ is a non-contractible loop.

\begin{remark}\label{rem:rot_jump_ind_cut}
The last integral in \eqref{eqn:rot_jump_ind} depends only on the loop $l_{\bfx'}$ and the locally compatible strain field $\bfveps$ prescribed on $\Omega$.
\end{remark}
\begin{figure}
\centering
\includegraphics[width=4.0in, height=3.0in]{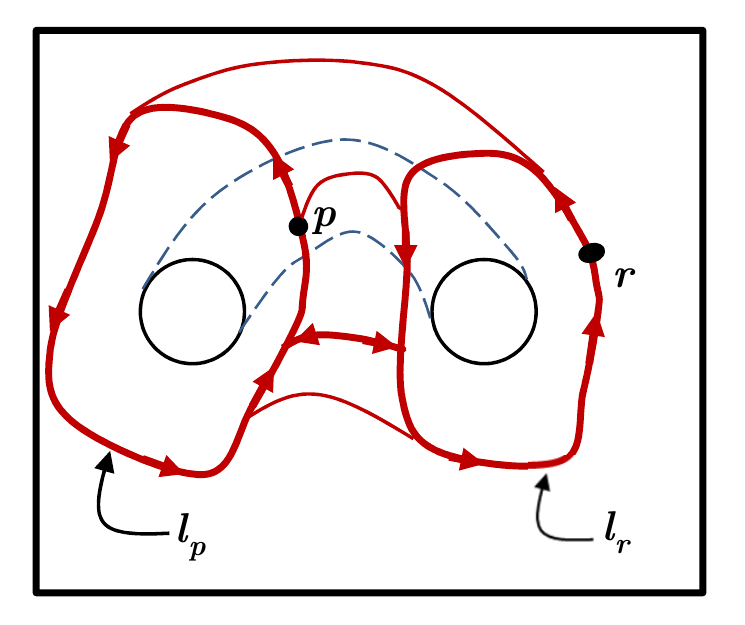}
\caption{A surface (with red outline) with two non-contractible loops (through $\bfp$ and $\bfr$) as edges. The dashed lines have the same meaning as in Fig. \ref{fig:fig_bodies}(a).}
\label{fig:fig_rot_ind_small}
\end{figure}

Consider two  mutually non-intersecting, non-contractible loops passing through points $\bfp, \bfr \in \Omega$ and connect them by a surface in $\Omega$ that has them as edges (see Figure \ref{fig:fig_rot_ind_small}). A closed curve can always be constructed on this surface that includes these loops as segments and two overlapping parts, as shown in Fig. \ref{fig:fig_rot_ind_small}. Applying Stokes theorem on this curve with integrand that of last integral in \eqref{eqn:rot_jump_ind}, and noting the local compatibility of $\bfveps$, we deduce that the loop integrals on $\bfl_{\bfp}$ and $\bfl_{\bfr}$ are equal, for the two points $\bfp, \bfr$ chosen arbitrarily with the constraint of non-intersection of the loops as mentioned. Let us denote this important fact as
\begin{equation}\label{eqn:rot_jump_ind_cut}
\int_{\bfl_{\bfx}} E_{ijm} (\bfs) \, ds_m = \int_{\bfl_{\bfx}} \left[ \veps_{im,j}(\bfs) - \veps_{jm,i} (\bfs) \right] \,ds_m =: \omega_{ij} \ \ \ \forall \bfx \in \Omega,
\end{equation}
where $\bfomega$ is a constant skew symmetric tensor on $\Omega$ and $\bfl_\bfx$ is any non-contractible loop through $\bfx$. Since this result \eqref{eqn:rot_jump_ind_cut} applies for all $\bfx \in \Omega$ without reference to any cut-surface, it applies to each point along the curve $\bfc_{(\bfx,\bfz)}$ under consideration in \eqref{eqn:pre_wein}, which by \eqref{eqn:rot_jump_ind} implies that 
\begin{lemma}\label{rem:rot_jump_ind_pos}
$\llbracket \lsup{\tau \omega_{ij}} \rrbracket$ in \eqref{eqn:pre_wein} is constant on $\tau$, and takes the same value on all cut-surfaces $\tau$ of $\Omega$.
\end{lemma}
Furthermore, the displacement jump across $\tau$ may be expressed as
\begin{equation}\label{eqn:wein}
\llbracket {^\tau \bfu} (\bfz) \rrbracket = \llbracket {^\tau \bfu} (\bfx) \rrbracket + \bfomega [\bfz - \bfx] \ \ \ \forall \bfz, \bfx \in \tau.
\end{equation}
Thus, thinking of $\bfx$ as fixed and $\bfz$ sweeping out $\tau$ and $^\tau \bfu^+$ and $^\tau \bfu^-$ as two displacements of the surface $\tau$, \eqref{eqn:wein} suggests that these two displacement fields of $\tau$ are related by an infinitesimally rigid deformation (cf. Definition \ref{def:rigid_small}). The statement \eqref{eqn:wein} is {\bf \emph{Weingarten's}} result \cite[as translated in \cite{delph_wein}]{wein} (see Nabarro \cite[pp 17-18]{nabarro} for another proof based on that in Love \cite{love}, but in more convenient notation).

\subsection{Volterra's ``characteristic of the distortion"}\label{sec:jump_rel_cut}
Given two cut-surfaces $\tau'$ and $\tau$ in $\Omega$  and two points $\bfx \in \tau'$ and $\bfy \in \tau$ that can be linked by a curve in $\Omega$ which intersects $\tau'$ and $\tau$ only at $\bfx$ and $\bfy$, respectively, we would now like to understand the relationship between the displacement jumps ${^{\tau'} \bfu}(\bfx)$ and ${^\tau \bfu} (\bfy)$. This question is motivated by statements in the classical literature starting from Volterra followed by Nabarro that the `infinitesimally rigid deformation' characterizing a Volterra dislocation is independent of the cut surfaces $\tau$ and $\tau'$. Before getting into the details, we first consider the statements from Volterra and Nabarro:
\begin{quote}
{\bf Volterra} \cite[Chapter II, as translated in \cite{delph_volt}]{volt} - ``\ldots if the multiply-connected elastic body is taken in its natural state then in order to bring it into a state of tension, one can perform the inverse operation - i.e., the sectioning that will render it simply connected - and then displace the two parts of each cut with respect to each other in such a manner that the relative displacements of the various pairs of pieces (which adhere to each other and which the cut has separated) are the resultants of translations and equal rotations; finally, re-establish the connectivity and the continuity along each cut, by subtracting or adding the necessary matter and welding the parts together. The set of these operations that relate to each cut may be called a \emph{distortion} of the body and the six constants may be called the \emph{characteristic of the distortion}.

\ldots One may say, in addition, that the six characteristics of each distortion are not elements that depend upon the location where the cut has been executed.

Indeed, that same process that served to establish formulas (III) for us proved that if one takes two cuts in the body then one may transform the one into the other by a continuous deformation, so the constants that relate to one cut are equal to the constants that relate to the other.

It then follows that the characteristics of a distortion are not elements that are specific to each cut, but they depend exclusively on the geometrical nature of the space that is occupied by the body and the regular deformation to which it has been subjected."
\end{quote}
The ``same process that served to establish formulas (III)" above refers to Volterra's proof of Weingarten's theorem, with the formulas stated as ``Upon denoting the six constants across each section by $l,m,n,p,q,r$, we have:
\[
(III) \ \ \ U = l + ry - qz, \ \ \ V = m + pz-rx, \ \ \ W = n + qx - py",
\]
where $U,V,W$ represent the three Cartesian components of the displacement jump.
\begin{quote}
{\bf Nabarro} \cite[p. 19]{nabarro} ``Volterra showed that any dislocation of Weingarten's type is equivalent to the dislocation produced by applying the same translation and rotation to the surfaces of any cut which can be continuously deformed into the original cut. This is proved by considering the body in its dislocated state, and showing that the same six constants $b_i$ and $d_{ij}$ are obtained by applying the preceding analysis to one cut or to the other." 

The ``preceding analysis" Nabarro refers to is the proof of Weingarten's theorem in his treatise \cite[pp 17-18]{nabarro} where $b_i$ refers to the the displacement jump at an arbitrarily fixed point of a cut and $d_{ij}$ refer to the components of the jump in the infinitesimal rotation tensor across the cut at that point.
\end{quote}
I was unable to find, or deduce (primarily due to my inability in forming a precise statement of the problem), a proof of these statements of Volterra and Nabarro. What I was able to deduce is a relationship between displacement jumps of `corresponding' points across two different cuts, with the sense of the correspondence defined in the opening paragraph of this Sec. \ref{sec:jump_rel_cut}. This is what is described in the following.
\begin{figure}
\centering
\includegraphics[width=3.5in, height=3.0in]{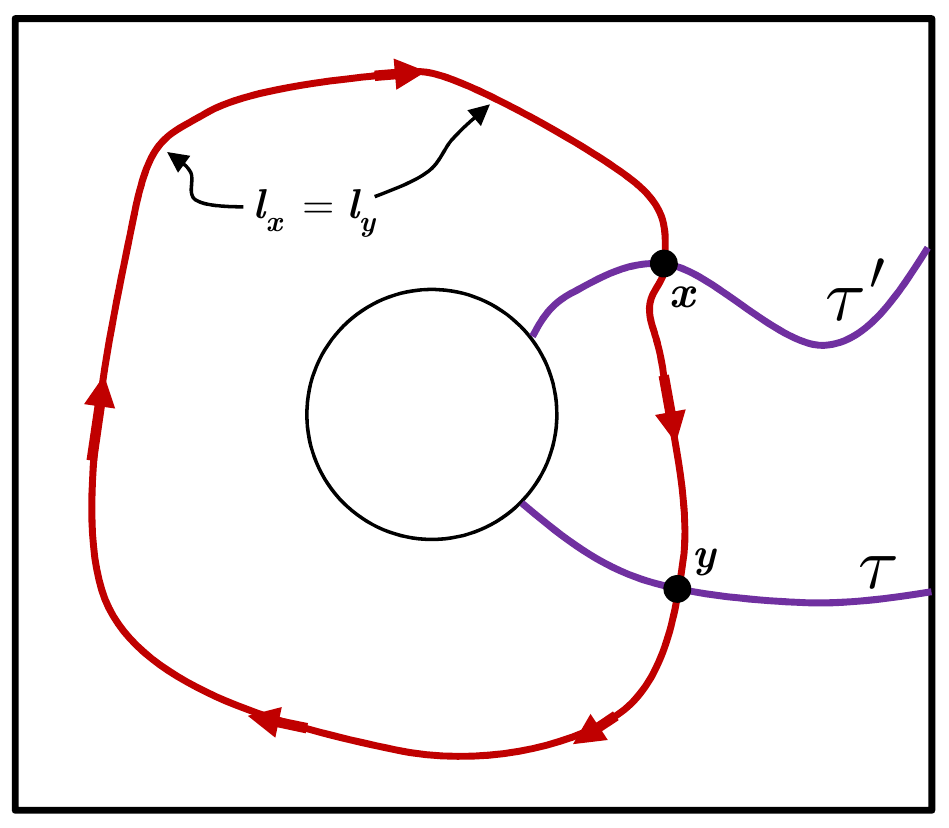}
\caption{A non-contractible loop intersecting cut-surfaces $\tau'$ and $\tau$ at points $\bfx$ and $\bfy$, respectively.}
\label{fig:two_cuts}
\end{figure}

With reference to Figure \ref{fig:two_cuts},
\begin{equation}\label{eqn:2cuts_1}
\lsup{{\tau'} u_i}(\bfx^+) - \lsup{{\tau'} u_i}(\bfx^-) = \int_{\bfc_{(\bfx^-,\bfx^+)}} \left( \veps_{ik}(\bfs) + \lsup{{\tau'} \omega_{ik}}(\bfs) \right) ds_k,
\end{equation}
where $\bfc_{(\bfx^-,\bfx^+)}$ is any curve in $\Omega$ starting at $\bfx^-$ and ending at $\bfx^+$. Thinking of the curve parametrized by $t \in [0,1]$ with $\bfs(0) = \bfx^-$ and $\bfs(1) = \bfx^+$ we have
\begin{equation*}
\begin{split}
\int_{\bfc(\bfx^-,\bfx^+)} \lsup{{\tau'} \omega_{ik}} (\bfs) \, ds_k & = \int^1_0 \lsup{{\tau'} \omega_{ik}} (\bfs(t)) \frac{ds_k}{dt} (t) \, dt\\
& = \lsup{{\tau'} \omega_{ik}}(\bfx^+) x^+_k - \lsup{{\tau'} \omega_{ik}}(\bfx^-) x^-_k - \int_{\bfc_{(\bfx^-,\bfx^+)}} s_j \,\lsup{{\tau'} \omega_{ij,k}} \, ds_k.
\end{split}
\end{equation*}
Taking the limit $\bfx^\pm \rightarrow \bfx$, and using \eqref{eqn:rot_jump_ind}, \eqref{eqn:rot_jump_ind_cut}, and \eqref{eqn:rotg_straing} in \eqref{eqn:2cuts_1} we obtain
\begin{equation}\label{eqn:loop_displ_jump_tau'}
\llbracket \lsup{{\tau'} u_i}(\bfx) \rrbracket =  \omega_{ik} x_k + \int_{\bfl_\bfx} \left[ \veps_{ik} (\bfs) - s_j (\veps_{ik,j} (\bfs) + \veps_{jk,i} (\bfs)) \right] \,ds_k,
\end{equation}
where $\bfl_\bfx$ represents a non-contractible loop passing through $\bfx$. Thus, the displacement jump at $\bfx \in \tau'$ depends on $\tau'$ only through $\bfx$, and otherwise on the constant $\bfomega$ on $\Omega$, and the line integral along a non-contractible loop passing through $\bfx$ of quantities depending only on the given strain field and the loop.

The displacement jump in ${^\tau \bfu}$ at $\bfy$ across $\tau$ may be expressed, following the same arguments to arrive at \eqref{eqn:loop_displ_jump_tau'}, as
\begin{equation}\label{eqn:loop_displ_jump_tau}
\llbracket \lsup{{\tau} u_i}(\bfy) \rrbracket =  \omega_{ik} y_k + \int_{\bfl_\bfy} \left[ \veps_{ik} (\bfs) - s_j (\veps_{ik,j} (\bfs) + \veps_{jk,i} (\bfs)) \right] \,ds_k.
\end{equation}
But, the hypothesis that $\bfx \in \tau'$ and $\bfy \in \tau$ can be linked by a curve means that we can always choose the non-contractible loop through $\bfx$ to pass through $\bfy \in \tau$ as well so that the choice of $\bfl_\bfx = \bfl_\bfy$ is admissible. Therefore, \eqref{eqn:loop_displ_jump_tau'} and \eqref{eqn:loop_displ_jump_tau} together imply that
\begin{equation}\label{eqn:displ_jump_tau_tau'}
\llbracket ^{\tau'} \bfu(\bfx) \rrbracket = \llbracket {^\tau} \bfu(\bfy) \rrbracket + \bfomega [\bfx - \bfy] \ \ \ \mbox{for} \ \ \bfx \in \tau' \ \mbox{and} \ \bfy \in \tau.
\end{equation}
When the surfaces $\tau$ and $\tau'$ can be mapped into each other by a continuous, 1-parameter family of surfaces (i.e., a homotopy), then $\bfx \in \tau'$ and $\bfy \in \tau$ can surely be linked by a curve. Thus for this situation, the displacement jumps across $\tau$ and $\tau'$ at corrresponding points that map into each other by the homotopy are related by \eqref{eqn:displ_jump_tau_tau'}.

We also note that, keeping $\bfx \in \tau'$ fixed and choosing two points $\bfy, \bfz \in \tau$ which can be linked to $\bfx \in \tau'$ by (different) curves, \eqref{eqn:displ_jump_tau_tau'} provides another proof of Weingarten's theorem \eqref{eqn:wein}.
\subsection{Burgers vector of a dislocation and its cut-surface independence}
While it is obvious that $\llbracket {^\tau \bfu} (\bfx) \rrbracket$  in \eqref{eqn:wein} is a constant translation vector for fixed $\bfx \in \tau$, it is clear that the choice of $\bfx$ is arbitrary and choosing some other base point changes this constant when $\bfomega \neq \bfzero$. Thus the rigid \emph{translation} in the Weingarten theorem \eqref{eqn:wein} is not well-defined when $\bfomega \neq \bfzero$. We note that given the function ${^\tau \bfu}$ and the family of all displacement fields on $\Omega_\tau$ related to it by infinitesimally rigid deformations, the jump $\llbracket {^\tau \bfu} (\bfx) \rrbracket$ for each fixed $\bfx \in \tau$ is unique within the family, see Remark \ref{rem:jump_family}. 
\begin{definition}
The Burgers vector of a Weingarten-Volterra defect of ${^\tau} \bfu$ is well-defined when $\llbracket {^\tau} \bfomega \rrbracket = \bfzero$ (in which case the defect is also called a dislocation), and is given by $\llbracket {^\tau} \bfu (\bfx) \rrbracket$ for \emph{any} $\bfx \in \tau$.
\end{definition}
\begin{remark}\label{rem:burgers_small_1}
The Weingarten result \eqref{eqn:wein} shows that the quantity 
\begin{equation}\label{eqn:fake_burgers}
\bfb_o^\tau := \llbracket {^\tau} \bfu (\bfx) \rrbracket - \bfomega \bfx
\end{equation}
 is a \emph{mathematical} constant for all $\bfx \in \tau$. This constancy however should not be interpreted as the \emph{physical} Burgers vector of the Weingarten-Volterra defect since it depends on the choice of the origin invoked to define position vectors  (see Remark \ref{rem:rigid_const} and cf. \cite{casey, zubov} - Zubov recognizes this problem \cite[p.19]{zubov}, but nevertheless adopts the definition \cite[Chapter 1.3]{zubov}). To see one of the problematic implications of such a definition, it is physically reasonable to expect that within the family of displacement fields of $\Omega_\tau$ that are related to each other by infinitesimally rigid deformations, the Burgers vector of a dislocation should, at most, be rotated and therefore maintain constant magnitude, and this is a physical statement where the choice of an origin to represent position vectors plays no role. It is easy to check that the expression $\bfb_0^\tau$ can be made to have arbitrary magnitude depending on the choice of origin when $\bfomega \neq \bfzero$.
\end{remark}
We now prove the following assertion:
\begin{theorem}\label{theo:burgers_small}
Given two cut-surfaces $\tau$ and $\tau'$ of $\Omega$, a locally compatible strain field $\bfveps$ on $\Omega$ with $\bfomega = \bfzero$ (defined by \eqref{eqn:rot_jump_ind_cut}), and two displacement fields ${^\tau} \bfu$ and $^{\tau'} \bfu$ compatible with $\bfveps_\tau$ on $\Omega_\tau$ and $\bfveps_{\tau'}$ on $\Omega_{\tau'}$, respectively, the Burgers vector, $\bfb$, of the dislocations of ${^\tau} \bfu$  and $^{\tau'} \bfu$ are equal and given by
\[
\bfb := \llbracket {^\tau} \bfu (\bfx) \rrbracket = \llbracket {^{\tau'}} \bfu (\bfy) \rrbracket, \ \  \bfx \in \tau, \bfy \in \tau'.
\]
\end{theorem}
\begin{remark}\label{rem:burgers_small_2}
In the existing literature, a distinction between the fields ${^\tau} \bfu$ and $^{\tau'} \bfu$ is generally not made and it is assumed without proof that $\llbracket {^\tau} \bfu (\bfx) \rrbracket = \llbracket ^{\tau'} \bfu (\bfx) \rrbracket$ when $\bfx \in \tau \cap \tau'$. We note that the domains of the two functions ${^\tau} \bfu:\Omega_{\tau} \rightarrow V_3$ and $^{\tau'} \bfu:\Omega_{\tau'} \rightarrow V_3$ are different, and it is not a priori obvious that the limiting value of the two displacement fields in question, or even their jumps, at any point of their corresponding cuts do not depend on the geometry of the respective cuts beyond the point of evaluation. In Proof 3 below, we fill this gap in the argument.
\end{remark}

We provide three different proofs of Theorem \ref{theo:burgers_small}, with different levels of assumptions.

\begin{figure}
\centering
\includegraphics[width=3.5in, height=3.0in]{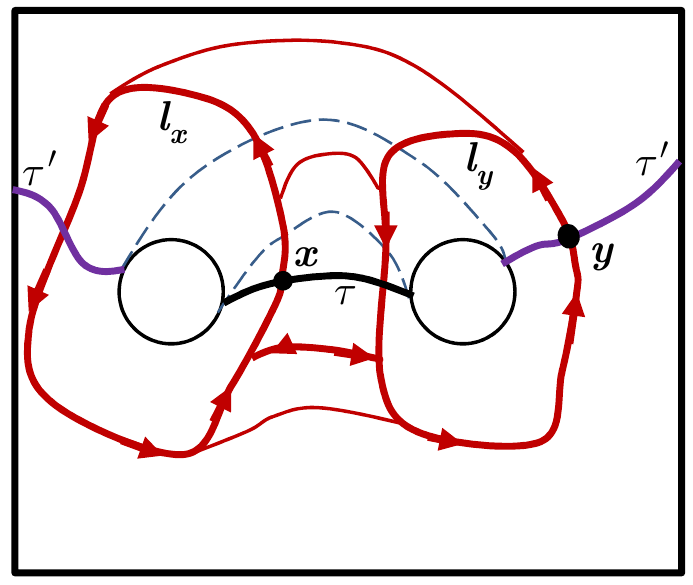}
\caption{Non-contractible loop $\bfl_\bfx$ intersects cut surface $\tau$ only once at $\bfx$. Non-contractible loop $\bfl_\bfy$ intersects cut surface $\tau'$ only once at $\bfy$.}
\label{fig:fig_disp_burgers_small}
\end{figure}

\emph{Proof 1}: For $\bfx_0 \in \Omega_\tau \cap \Omega_{\tau'}$, from \eqref{eqn:rotg_straing} we have that any $^\tau \bfomega$ and $^{\tau'} \bfomega$ may be expressed as
\begin{equation}\label{eqn:rot_rep}
\begin{split}
& {^\tau \omega_{ij}} (\bfz) = {^\tau \omega_{ij}} (\bfx_0) + \int_{\bfc_{(\bfx_0,\bfz)}} E_{ijk} (\bfs) \, ds_k \ \ \mbox{for} \ \  \bfz \in \Omega_\tau,\\
&{^{\tau'} \omega_{ij}} (\bfy) = {^{\tau'} \omega_{ij}} (\bfx_0) + \int_{\bfc'_{(\bfx_0,\bfy)}} E_{ijk} (\bfs) \, ds_k \ \ \mbox{for} \ \  \bfy \in \Omega_{\tau'},
\end{split}
\end{equation}
where $\bfc$ is a curve in $\Omega_\tau$ and $\bfc'$ is a curve in $\Omega_{\tau'}$. Then for $\bfx \in \tau$ we have that $\llbracket {^\tau} \omega_{ij} (\bfx) \rrbracket = \int_{l_{\bfx}} E_{ijk} (\bfs) \, ds_k  = \omega_{ij} = 0$ from  \eqref{eqn:rot_jump_ind} and \eqref{eqn:rot_jump_ind_cut}, and similarly, for $\bfx \in \tau'$, $\llbracket {^{\tau'}} \omega_{ij} (\bfx) \rrbracket = 0$. This implies that both $^\tau \bfomega$ and $^{\tau'} \bfomega$ are actually continuous fields on $\Omega$. We now have that both representations in \eqref{eqn:rot_rep} are actually valid on $\Omega$ and that
\[
{^{\tau'} \bfomega} (\bfx) = {^{\tau} \bfomega} (\bfx) + \bfA \ \ \forall \bfx \in \Omega,
\]
where $\bfA$ is a constant  skew-symmetric second order tensor on $\Omega$. From \eqref{eqn:rotg_straing} and the above relation we also have that
\begin{equation*}
\begin{split}
& ^{\tau} u_{i,j} (\bfx) = {^{\tau}} \omega_{ij} (\bfx) + \veps_{ij} (\bfx) \ \ \forall \bfx \in \Omega_\tau,\\
& {^{\tau'}} u_{i,j} (\bfx) = {^{\tau}} \omega_{ij} (\bfx) + A_{ij} + \veps_{ij} (\bfx) \ \ \forall \bfx \in \Omega_{\tau'}, 
\end{split}
\end{equation*}
so that for $\bfx \in \tau$ and $\bfy \in \tau'$ we have
\begin{equation}\label{eqn:jump_utau_utau'}
\llbracket {^\tau} \bfu (\bfx) \rrbracket = \int_{\bfl_\bfx} [ {^\tau} \bfomega(\bfs) + \bfveps(\bfs) ]\, d\bfs; \ \ \ \ \llbracket {^{\tau'}} \bfu (\bfy) \rrbracket = \int_{\bfl_\bfy} [ {^\tau} \bfomega(\bfs) + \bfveps(\bfs) ]\, d\bfs,
\end{equation}
where $\bfl_\bfx$ and $\bfl_\bfy$ are non-contractible loops through $\bfx$ and $\bfy$ that intersect $\tau$ and $\tau'$ exactly once, respectively. But ${^\tau} \bfomega + \bfveps =: \bfH$ is a continuous, and twice differentiable field on $\Omega$ that satisfies $H_{ik,l} = H_{il,k}$ on it by using $^\tau \omega_{ij,k} = E_{ijk}$. Then, by Stokes' theorem, we have that the two loop integrals in \eqref{eqn:jump_utau_utau'} are equal (by connecting them by a surface in $\Omega$ with the loops as boundaries and considering a closed curve on this surface with these loops as segments along with two overlapping parts, as in Fig. \ref{fig:fig_disp_burgers_small}), and the proof is complete.

\emph{Proof 2}: We assume that there is at least one point $\bfx \in \tau'$ and $\bfy \in \tau$ which can be connected by a curve in $\Omega$. Then, consider \eqref{eqn:displ_jump_tau_tau'} for $\bfomega = \bfzero$. Finally, we apply Weingarten's theorem \eqref{eqn:wein} to each surface. The proof is complete.

\emph{Proof 3}: Assume that it is always possible to join $\tau$ and $\tau'$ by another cut-surface $\tau''$ such that $\tau''$ has common parts with $\tau$ and $\tau'$. Let $\bfx \in \tau \cap \tau''$ and $\bfy \in \tau' \cap \tau''$. We also assume that it is possible to choose $\tau''$ and a non-contractible loop $\bfl_\bfx$ such that the loop intersects $\tau$ and $\tau''$ exactly once at $\bfx$ and, similarly, a choice of $\bfl_\bfy$ can be made w.r.t $\tau'$ and $\tau''$, intersecting both exactly once at $\bfy$. Then, \emph{from the arguments leading up to \eqref{eqn:loop_displ_jump_tau'},} we conclude that $\llbracket {^\tau} \bfu (\bfx) \rrbracket = \llbracket {^{\tau''}} \bfu (\bfx) \rrbracket$ and $\llbracket {^{\tau'}} \bfu (\bfy) \rrbracket = \llbracket {^{\tau''}} \bfu (\bfy) \rrbracket$. But then, Weingarten's theorem \eqref{eqn:wein} for $\bfomega = \bfzero$ applied to all three surfaces  completes the proof.

\section{Finite deformation}\label{sec:large}
The primary difference between the concepts and methods employed in proving Weingarten's theorem and associated results between the settings of infinitesimal and finite deformation kinematics is that the right hand side of the equation of integrability  \eqref{eqn:pfaff_large}$_2$ is not specified in terms of given data unlike \eqref{eqn:rotg_straing}$_3$. This takes the great power afforded by Stokes' theorem essentially out of play in the case of discussing integrability for finite kinematics, using similar techniques as for the results of small deformation theory.

\subsection{Weingarten's theorem for finite deformation}
Given a simply connected $\Omega_\tau$ (induced from the non-simply connected $\Omega$ with a cylindrical/toroidal hole) and a locally compatible $\bfC$ field on it, consider any deformation ${^\tau\bfy}$ compatible with $\bfC_\tau$ on $\Omega_\tau$.  Appendix \ref{comp_large} shows that that there exists a family of such deformation fields, each member of which satisfies
\begin{equation}\label{eqn:pfaff_large}
\begin{split}
^\tau y^i_{,\alpha} & = {^\tau F^i}_\alpha,\\
{^\tau F^i}_{\alpha,\beta} & = \Gamma^\rho_{\alpha\beta}\, {^\tau F^i}_\rho,
\end{split}
\end{equation}
where $\Gamma$ is defined in \eqref{eqn:loc_comp}. To be precise, the pair $({^\tau \bfy}, {^\tau \bfF})$ for fixed $\tau$ is not unique, but we do not make this distinction explicit in notation to keep it manageable (unless absolutely essential).

As in Section \ref{sec:wein_small}, consider a curve $\bfc_{(\bfx,\bfz)}$ on the surface $\tau$ joining $\bfx$ and $\bfz$. Corresponding to $\bfc$, consider two other curves $\bfc^+_{(\bfx^+,\bfz^+)}$ and $\bfc^-_{(\bfx^-,\bfz^-)}$ in $\Omega_\tau$ on either side of, and close to, $\tau$. The curves run from $\bfx^\pm$ to $\bfz^\pm$ (in obvious notation), and in the following we will be thinking of limits of product integrals (see \cite[Section 1.1]{dol_fried} for definition) along $\bfc^\pm$ as $\bfc^\pm$ tend to $\bfc$, with both $\bfc^\pm$ parametrized by $s \in [0,1]$. 
\begin{definition}\label{def:G_P}
Given a parametrized curve of position vectors $\bfa: [x,y] \rightarrow V_3, [x,y] \subset \R$, the $3 \times 3$ matrix-valued function of the parameter of the curve, $G^\bfa(s)$ is defined as
\[
{(G^\bfa)}^{\rho}_{\alpha} (s) := \Gamma^\rho_{\alpha \beta} (\bfa(s))\, \frac{da^ \beta}{ds}(s);
\]
the $3 \times 3$ invertible-matrix ${\sf{P}}^\bfa$ is defined by the product integral \cite{dol_fried}
\[
{\sf{P}}^\bfa = \prod^{y}_{x} e^{{G}^{\bfa} (s) ds}.
\]
In standard matrix notation $G^\rho_\alpha$ and $\left({\sf{P}}^\bfa\right)^\rho_\alpha$ are to be interpreted as $G_{\rho \alpha}$ and $\left({\sf{P}}^\bfa\right)_{\rho \alpha}$ and our definition of the product integral corresponding to the matrix $G$ is identical to the definition of the product integral corresponding to the matrix $G^T$ given in \cite[Theorem 1.1]{dol_fried}.
\end{definition}

Along such curves we have
\begin{equation}\label{eqn:def_G}
\begin{split}
 & {^\tau F^i_{\alpha,\beta}}(\bfc^\pm(s))\,\frac{d(c^\pm)^ \beta}{ds}(s) = \Gamma^\rho_{\alpha \beta} (\bfc^\pm(s))\, \frac{d(c^\pm)^ \beta}{ds}(s)\, {^\tau F^i_{\rho}}(\bfc^\pm(s))\\
& \Longrightarrow \frac{d ({^\tau F^i_\alpha}\circ \bfc^\pm)}{ds}(s)  = {(G^{\bfc^\pm})}^\rho_\alpha (s)\, {^\tau F^i_{\rho}}(\bfc^\pm(s)).
\end{split}
\end{equation}
Equation \eqref{eqn:def_G}$_2$ implies that ${^\tau F^i_{\alpha}}(\bfz^\pm)$ has the representation shown below in \eqref{eqn:def_P} by the representation and uniqueness of solutions of linear systems of ordinary differential equations with prescribed initial data proved in \cite[Theorem 2.1, pp 12-13]{dol_fried} (of course, uniqueness can also be proved by the Gronwall inequality). Thus,
\begin{equation}\label{eqn:def_P}
\begin{split}
{^\tau F^i_{\alpha}}(\bfz^\pm) = \left[ \prod^{s=1}_{s=0} e^{{G}^{\bfc^\pm} (s) ds} \right] ^\rho_\alpha {^\tau F^i_{\rho}}(\bfx^\pm) \Longleftrightarrow {^\tau \bfF} (\bfz^\pm) = {^\tau \bfF} (\bfx^\pm) {\sf{P}}^{\bfc^\pm},
\end{split}
\end{equation}
 so that
\begin{equation*}
{^\tau \bfF} (\bfz^+) \left[ {^\tau \bfF} (\bfz^-)\right]^{-1} = {^\tau \bfF} (\bfx^+) {\sf{P}}^{\bfc^+} \left[ {\sf{P}}^{\bfc^-} \right]^{-1}  \left[ {^\tau \bfF} (\bfx^-) \right]^{-1}.
\end{equation*}
Then , in the limit $\bfc^\pm \rightarrow \bfc$, we have ${\sf{P}}^{\bfc^+} = {\sf{P}}^{\bfc^-} = {\sf{P}}^{\bfc}$ and therefore
\begin{equation}\label{eqn:pre_wein_large}
{^\tau \bfQ}:= {^\tau \bfF}^+ (\bfz) \left[ {^\tau \bfF}^- (\bfz)\right]^{-1} = {^\tau \bfF}^+ (\bfx) \left[ {^\tau \bfF}^- (\bfx)\right]^{-1} \ \ \ \forall \, \bfx, \bfz \in \tau.
\end{equation}
The compatibility of $^\tau \bfy$ with $\bfC_\tau$ implies that $^\tau \bfF$ is invertible. Since $\bfC =: \bfU^2$ is continuous on $\Omega$, ${^\tau \bfF}^\pm = {^\tau \bfR}^\pm \bfU$ on $\tau$ with ${^\tau \bfR}^\pm$  proper orthogonal, and hence ${^\tau \bfQ} =  {^\tau \bfR}^+ \left[ {^\tau \bfR}^- \right]^T $ is a proper orthogonal tensor.

Considering now the first equation in \eqref{eqn:pfaff_large} we have
\begin{equation*}
\begin{split}
& {^\tau y^i_{,\alpha}} (\bfc^\pm(s)) \frac{d(c^\pm)^\alpha}{ds}(s) = {^\tau F^i_\alpha} (c^\pm(s))\frac{d(c^\pm)^\alpha}{ds}(s)\\
 \Longrightarrow & {^\tau \bfy^\pm}(\bfz) - {^\tau \bfy^\pm}(\bfx) = \lim_{\bfc^\pm \rightarrow \bfc} {^\tau \bfy}(\bfz^\pm) - {^\tau \bfy}(\bfx^\pm) = \lim_{\bfc^\pm \rightarrow \bfc} \int_0^1 {^\tau \bfF}(\bfc^\pm(s)) \frac{d \bfc^\pm}{ds} (s) \,ds.
\end{split}
\end{equation*}
Using \eqref{eqn:pre_wein_large} we obtain
\begin{equation}\label{eqn:wein_large}
\begin{split}
{^\tau \bfy^+}(\bfz) - {^\tau \bfy^+}(\bfx) & = \int_0^1 {^\tau \bfF^+}(\bfc(s)) \frac{d \bfc}{ds} (s) \,ds \\
& =  {^\tau \bfQ}\int_0^1 {^\tau \bfF^-}(\bfc(s)) \frac{d \bfc}{ds} (s) \,ds = {^\tau \bfQ} \left( \lim_{\bfc^- \rightarrow \bfc} \int_0^1 {^\tau \bfF}(\bfc^-(s)) \frac{d \bfc^-}{ds} (s) \,ds \right)\\
& =  {^\tau \bfQ} \left( \lim_{\bfc^- \rightarrow \bfc} \left[ {^\tau \bfy}(\bfz^-) - {^\tau \bfy}(\bfx^-) \right]  \right)\\
& = {^\tau \bfQ} \left[ {^\tau \bfy^-}(\bfz) - {^\tau \bfy^-}(\bfx) \right] \ \ \ \forall \, \bfz, \bfx \in \tau.
\end{split}
\end{equation}
Thus, thinking of  $^\tau \bfy^+$ and $^\tau \bfy^-$ as two deformations of the surface $\tau$, \eqref{eqn:wein_large} suggests that these two deformation fields of $\tau$ are related by a rigid deformation (cf. Definition \ref{def:rigid_large}).
\emph{This is a proof of Weingarten's theorem at finite deformation, essentially due to} {\bf \emph{Zubov}} \cite[Sec. 1.3]{zubov}(also see {\bf \emph{Casey}} \cite{casey} for a different proof).

\begin{definition}\label{def:discl_strength}
${{^{^\tau \bfy}} \bfQ}$, or $^\tau \bfQ$ when there is no ambiguity in the rotation of which deformation is being referred to, $(\bfomega)$ is defined to be the disclination strength of the Weingarten-Volterra defect of ${^\tau \bfy}$ $(^\tau \bfu)$.
\end{definition}
\begin{remark}\label{rem:RQRt}
Let $^\tau \bfy_1$ and $^\tau \bfy_2$ be two deformations compatible with $\bfC_\tau$ on $\Omega_\tau$, with deformation gradient fields ${^\tau \bfF_2}$ and ${^\tau \bfF_1}$. Then, necessarily, there exists a constant orthogonal tensor $\bfR^*$ on $\Omega_\tau$ such that ${^\tau \bfF_2} = \bfR^* \, {^\tau \bfF_1}$ and therefore ${^\tau \bfQ}_2 = \bfR^* \, {^\tau \bfQ}_1 \, {\bfR^*}^T $. Thus, in contrast with the small deformation case (see Remark \ref{rem:jump_family}), the disclination strength of the Weingarten-Volterra defect of $^\tau\bfy$, ${^{^\tau \bfy}} \bfQ$, is not constant for all  deformations $^\tau \bfy$ of $\Omega_\tau$ compatible with $\bfC_\tau$ that are rigidly related to each other.

Zubov \cite[p. 20]{zubov} claims that his ``vector of finite rotation" is uniquely determined by the field $\bfC$ (in our notation) for a doubly-connected domain in nonlinear elasticity, which implies from \cite[Equation (1.3.5)]{zubov} that $^\tau \bfQ$ must be too. The demonstration above shows that this is not the case.\footnote{
As an aside, Zubov's notation is non-standard, e.g., the action of a tensor $\bfA$ on a vector $\bfb$ is written as $\bfb \cdot \bfA$; for $Q^M$ (curvilinear) coordinates on the current configuration  with position vectors represented as $\bfR$ and $q_s$ as coordinates on the reference configuration with position vectors represented as $\bfr$, the deformation gradient is written as $\left(\frac{\partial Q^M}{\partial q^s}\right) \bfr^s \otimes \bfR_M$, where $\bfr^s$ represents (an element of) the dual basis in the reference configuration corresponding to coordinates $q_s$, and $\bfR_M$ is the natural basis in the current configuration (instead of the more standard notation that would be $\left(\frac{\partial Q^M}{\partial q^s}\right) \bfR_M \otimes \bfr^s$; the correspondence of upper and lower case letters with objects on the current and reference configuration in this footnote also follows Zubov's notation).}
\end{remark}

\subsection{A condition for $\tau-$independence of ${^\tau \bfQ}$}\label{sec:Q_tau_ind}
With reference to Remark \ref{rem:burgers_small_2}, Zubov \cite[p. 19]{zubov} does not make a distinction between the fields ${^\tau} \bfy: \Omega_\tau \rightarrow V_3$ and $^{\tau'} \bfy: \Omega_{\tau'} \rightarrow V_3$ compatible with $\bfC_\tau$ and $\bfC_{\tau'}$, respectively, for two different cut-surfaces $\tau$ and $\tau'$. He also assumes that ${^\tau \bfQ} (\bfx) = {^{\tau'} \bfQ} (\bfx)$ if $\bfx \in \tau \cap \tau'$ without proof. Clearly, Remark \ref{rem:RQRt} suggests that this need not be true without further conditions, even when $\tau = \tau'$. In this section we define a sufficient condition that ensures the cut-surface independence of $^\tau \bfQ$.

\begin{figure}
\centering
\includegraphics[width=4.0in, height=3.5in]{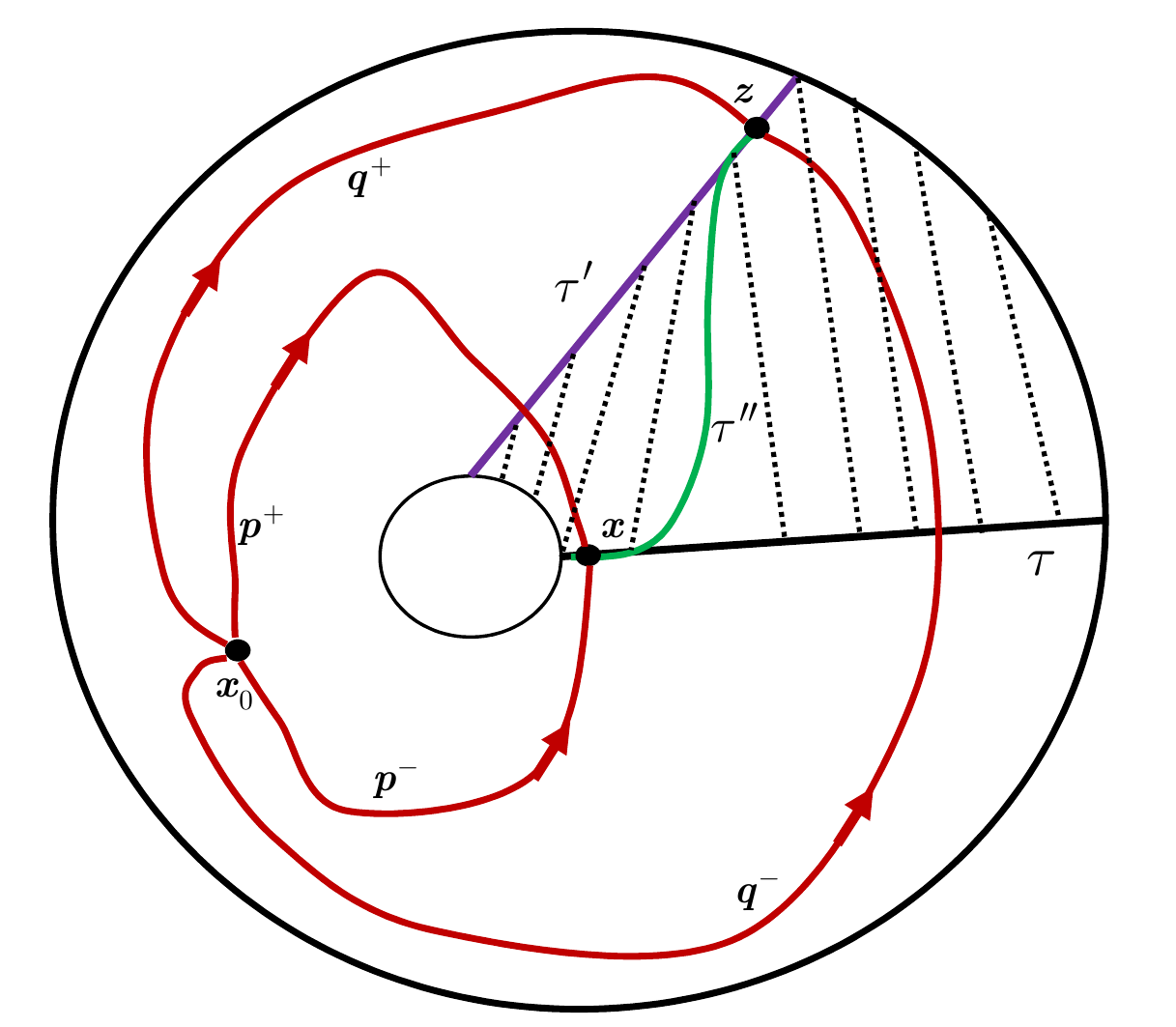}
\caption{$\bfx_0$ placed in the shaded region would not satisfy the hypotheses of the argument in Sec. \ref{sec:Q_tau_ind}.}
\label{fig:Q_ind}
\end{figure}

Let $\tau$, $\tau'$, and $\tau''$ be cut-surfaces that render $\Omega$ simply connected. Let $\tau \cap \tau''$ and $\tau' \cap \tau''$ both be non-empty. Assume there exist a point $\bfx_0 \in \Omega$ and an $\bfx \in \tau \cap \tau''$ such that they can be connected by two paths $\bfp^+$ and $\bfp^-$, both contained in $\Omega_\tau \cup \{\bfx\}$ and in $\Omega_{\tau''} \cup \{\bfx\}$, where $\bfp^+$ and $\bfp^-$ approach $\bfx$ from opposite sides of $\tau$ and $\tau''$. Similarly, assume that $\bfx_0$ can be connected to a $\bfz \in \tau' \cap \tau''$ by two paths $\bfq^+$ an $\bfq^-$, both contained in $\Omega_\tau' \cup \{\bfz\}$ and in $\Omega_{\tau''} \cup \{\bfz\}$, where $\bfq^+$ and $\bfq^-$ approach $\bfz$ from opposite sides of $\tau'$ and $\tau''$ - see Figure. \ref{fig:Q_ind} for a realization of these conditions. Consider deformations $^\tau \bfy: \Omega_\tau \rightarrow V_3$, $^{\tau'} \bfy: \Omega_{\tau'} \rightarrow V_3$  and $^{\tau''} \bfy: \Omega_{\tau''} \rightarrow V_3$ with deformation gradient fields ${^\tau \bfF}$, ${^{\tau'}\bfF}$, and ${^{\tau''}\bfF}$, respectively, with 
\begin{equation}\label{eqn:F_ic}
^\tau \bfF(\bfx_0) = {^{\tau'} \bfF}(\bfx_0) =  {^{\tau''}\bfF}(\bfx_0) = \bfF_0.
\end{equation}
Under these hypotheses, arguments identical to arriving at \eqref{eqn:def_G}, \eqref{eqn:def_P}, and \eqref{eqn:pre_wein_large} imply
\begin{equation*}
\begin{split}
& {^\tau \bfF^\pm}(\bfx) = \bfF_0 {\sf P}^{\bfp^\pm} = {^{\tau''} \bfF^\pm}(\bfx),\\
& {^{\tau'} \bfF^\pm}(\bfz) = \bfF_0 {\sf P}^{\bfq^\pm} = {^{\tau''} \bfF^\pm}(\bfz),
\end{split}
\end{equation*}
and using \eqref{eqn:pre_wein_large} we obtain
\begin{equation}\label{eqn:Q_ind_large}
{^\tau \bfQ}(\bfx) = {^{\tau'} \bfQ}(\bfy) = {^{\tau''} \bfQ}(\bfz) \ \ \ \forall \ \bfx \in \tau, \forall \ \bfy \in \tau', \forall \ \bfz \in \tau'',
\end{equation}
\emph{for the ${^{(\cdot)}} \bfQ$ generated from deformations ${^{(\cdot)}} \bfy$ that satisfy \eqref{eqn:F_ic} on their respective regions $\Omega_{(\cdot)}$, with $\bfx_0$ and the cut-surfaces, $\tau$, $\tau'$, and $\tau''$ satisfying the hypotheses mentioned above} in the second paragraph of this section \ref{sec:Q_tau_ind}.
\begin{remark}
Related to the discussion in \cite[Sec. 1.3]{zubov} of the vector of finite rotation, Zubov mentions `initial conditions' \cite[Equation (1.2.5), Sec. 1.2]{zubov} specifiable at any point of $\Omega$.  The point $\bfx_0$ where `initial conditions' \eqref{eqn:F_ic} may be specified for the proof above requires further specification of  a topological nature (see Fig. \ref{fig:Q_ind}).
\end{remark}
\begin{remark}
Erroneous remarks about the results \eqref{eqn:wein_large} and \eqref{eqn:Q_ind_large} are made in a footnote in \cite[p.146]{AF15}. While the footnote does not affect the developments of \cite{AF15} in any way, the latter dealing with a different, but broadly related, geometric construct for line defects than that of Weingarten-Volterra, nevertheless, the error is entirely regretted.
\end{remark}
\begin{remark}
The contrast in the generality of the result \eqref{eqn:Q_ind_large} for large deformations and the corresponding result, Remark \ref{rem:rot_jump_ind_pos}, for small deformations is to be noted. In particular, the small deformation result requires no specification of `initial conditions.'
\end{remark}

\subsection{Burgers vector of a dislocation and its dependence on the cut-surface}\label{subsec:burgers_large}
It is clear from \eqref{eqn:wein_large} that when ${^\tau \bfQ} = \bfI$, the jump in $^\tau \bfy$ across $\tau$ is constant on $\tau$ as  well as being a well-defined physical constant independent of the origin, unlike ${^\tau \bfy^+} - {^\tau \bfQ} \, {^\tau \bfy^-}$ which is also a constant on $\tau$ but not independent of the choice of origin (cf. \cite[p. 481]{casey}, \cite[p. 19]{zubov}). Hence, we define
\begin{definition}
The Burgers vector of a Weingarten-Volterra defect of ${^\tau} \bfy$ exists when ${^\tau} \bfQ = \bfI$ (in which case the defect is also called a dislocation), and is given by $\llbracket {^\tau} \bfy (\bfx) \rrbracket$ for \emph{any} $\bfx \in \tau$.
\end{definition}
\begin{remark}
Let $^\tau \bfy_1$ and $^\tau \bfy_2$ be two deformations compatible with $\bfC_\tau$ on $\Omega_\tau$. Then, necessarily, there exists a constant orthogonal tensor $\bfR^*$ on $\Omega_\tau$ such that ${^\tau \bfy_2} = \bfR^* \, {^\tau \bfy_1} + \bft$, with $\bft$ a constant on $\Omega_\tau$ and therefore $\llbracket {^\tau \bfy_2} \rrbracket = \bfR^* \, \llbracket {^\tau \bfy_1} \rrbracket$ on $\tau$, both jumps not necessarily constant on the cut-surface $\tau$.

When ${^{^\tau \bfy_1} \bfQ} = \bfI \Longleftrightarrow {^{^\tau \bfy_2} \bfQ} = \bfI$ by Remark \ref{rem:RQRt}, $\llbracket {^\tau \bfy_1} \rrbracket$ and $\llbracket {^\tau \bfy_2}\rrbracket$ are the constant Burgers vectors of the two deformations $^\tau \bfy_1$ and  $^\tau \bfy_2$, respectively, and are related by the rotation $\bfR^*$ linking the two deformations; thus they are not uniquely determined by the $\bfC$ field on $\Omega$ (cf. \cite[p. 20]{zubov}).
\end{remark}
We prove the following assertion:
\begin{theorem}\label{theo:burgers}
Consider two cut-surfaces $\tau$ and $\tau'$ of $\Omega$, a locally compatible strain field $\bfC$ on $\Omega$, and two deformations ${^\tau} \bfy$ and $^{\tau'} \bfy$ compatible with $\bfC_\tau$ on $\Omega_\tau$ and with $\bfC_{\tau'}$ on $\Omega_{\tau'}$, respectively, satisfying ${^\tau \bfQ} = \bfI$ and ${^{\tau'} \bfQ} = \bfI$. Suppose there exists a non-contractible loop $\bfl_\bfx \subset \Omega$ passing through $\bfx \in \Omega$ that intersects both $\tau$ and $\tau'$ exactly once. Then the Burgers vector of the dislocations of ${^\tau} \bfy$  and $^{\tau'} \bfy$ are related by 
\begin{equation}\label{eqn:burg_large}
\left[^\tau \bfR (\bfx) \right]^T \left\llbracket {^\tau} \bfy (\bfs) \right\rrbracket = \left[^{\tau'} \bfR (\bfx) \right]^T \left\llbracket {^{\tau'} \bfy} (\bfz) \right\rrbracket , \ \  \bfs \in \tau, \bfz \in \tau', 
\end{equation}
where $^\tau \bfR$ and $^{\tau'}\bfR$ are the rotation tensor fields from the polar decompositions of $^\tau \bfF = {^\tau\bfR}\, \bfU$ and ${^{\tau'} \bfF} = {^{\tau'} \bfR} \, \bfU$, respectively. In particular, the magnitudes of the two Burgers vectors are equal.

Furthermore, if the values of ${^\tau \bfF} (\bfx^*) = {^{\tau'} \bfF} (\bfx^*) = \bfF_0$ for some $\bfx^* \in \Omega$, then the Burgers vectors of the dislocations of ${^\tau} \bfy$ and $^{\tau'} \bfy$ are equal.
\end{theorem}

\emph{Proof:} Since $^\tau \bfQ = {^{\tau'} \bfQ} = \bfI$, $^\tau \bfF$ and ${^{\tau'}} \bfF$, solutions of \eqref{eqn:pfaff_large}$_2$ on $\Omega_\tau$ and $\Omega_{\tau'}$, respectively, are actually continuous functions on $\Omega$; for the same reason, it is clear from \eqref{eqn:wein_large} that it suffices to prove the theorem statement for one $\bfs \in \tau$ and one $\bfz \in \tau'$.

 It is also clear that
\[
\left\llbracket {^\tau} \bfy (\bfs^*) \right\rrbracket = \int_{\bfl_\bfx} {^\tau \bfF} \circ \bfl(t) \frac{d\bfl}{dt}(t) \,dt; \ \ \ \ \left\llbracket {^{\tau'} \bfy} (\bfz^*) \right\rrbracket = \int_{\bfl_\bfx} {^{\tau'} \bfF} \circ \bfl(t) \frac{d\bfl}{dt}(t) \,dt,
\]
where $\bfs^*$ is the point at which $\bfl_\bfx$ intersects $\tau$ and $\bfz^*$ is the point at which $\bfl_\bfx$ intersects $\tau'$. 

We think of the closed loop $\bfl_\bfx$ as parametrized by $t \in [0,1]$ with $\bfl(0) = \bfl(1) = \bfx$ and denote the portion of the curve corresponding to the interval $[0,t], t \leq 1$ as $\bfl_t$. Then
\[
{^\tau \bfF} \circ \bfl (t) = {^\tau \bfF}(\bfx) {\sf{P}}^{\bfl_t}; \ \ \ \ {^{\tau'} \bfF} \circ \bfl (t) = {^{\tau'} \bfF}(\bfx) {\sf{P}}^{\bfl_t}
\]
so that
\[
{^\tau \bfF} \circ \bfl (t) = {^\tau \bfF}(\bfx) \left[ {^{\tau'} \bfF}(\bfx) \right]^{-1} {^{\tau'} \bfF} \circ \bfl (t) = {^\tau \bfR}(\bfx) \left[ {^{\tau'} \bfR}(\bfx) \right]^{T} \, {^{\tau'}} \bfF \circ \bfl (t),
\]
and we have the desired result \eqref{eqn:burg_large} by performing a line integral of both sides of the expression along the loop $\bfl_\bfx$. 

Since $^\tau \bfF$ and $^{\tau'} \bfF$ are both continuous functions on $\Omega$ and satisfy
\[
\left[ {^\tau \bfF}(\bfr) - {^{\tau'} \bfF(\bfr)} \right] = \left[ {^\tau \bfF}(\bfx^*) - {^{\tau'} \bfF(\bfx^*) } \right]{\sf{P}}^{\bfc},
\]
where $\bfc$ is a parametrized curve from $\bfx^*$ to each $\bfr \in \Omega$, then, whenever ${^\tau \bfF} (\bfx^*) = {^{\tau'} \bfF} (\bfx^*) = \bfF_0$, we have uniqueness, i.e., ${^\tau \bfF} = {^{\tau'} \bfF}$ on $\Omega$, which further implies that ${^\tau \bfR} (\bfx) = {^{\tau'} \bfR (\bfx)}$ and hence the Burgers vector of $^\tau \bfy$ and $^{\tau'} \bfy$ are equal.
\begin{remark}
Let $\bfl_\bfz \subset \Omega$ also be a non-contractible loop  passing through $\bfz \in \Omega$ that intersects both $\tau$ and $\tau'$ exactly once. Denoting the (constant on $\tau$) Burgers vector of $^\tau \bfy$ as $^\tau\bfb$ and the Burgers vector for $^{\tau'} \bfy$ as $^{\tau'}\bfb$, \eqref{eqn:burg_large} implies
\[
\left[^\tau \bfR (\bfx) - {^\tau \bfR} (\bfz) \right]^T {^\tau \bfb} = \left[^{\tau'} \bfR (\bfx) - {^{\tau'} \bfR} (\bfz) \right]^T {^{\tau'} \bfb}.
\]
Thus, the change in the action of the (transposed) rotation of $^\tau \bfy$ on the latter's Burgers vector in moving from $\bfx \in \Omega$ to $\bfz \in \Omega$ is equal to the change in the action of the (transposed) rotation of $^{\tau'} \bfy$ on its Burgers vector for the same movement in point of evaluation. 
\end{remark}

While it is natural in the context of dislocations to assume that $^\tau \bfQ = \bfI$ and $^{\tau'} \bfQ = \bfI$, we now consider an argument which shows that assuming only one of these conditions implies the other in many circumstances, without invoking any `initial conditions' of the type \eqref{eqn:F_ic}, Sec. \ref{sec:Q_tau_ind}.

Let $\bfx \in \tau$ and $\bfz \in \tau'$. Assume that a cut-surface ${\tau''}$ exists that contains both $\bfx$ and $\bfz$.  Consider points $\bfx^+$ and $\bfx^-$ on opposite sides of $\tau''$ and near it. Consider curves $\bfl_{(\bfx^-,\bfx^+)}$ and $\bfq_{(\bfx^+,\bfx^-)}$ in $\Omega_{\tau''}$ connecting $\bfx^-$ to $\bfx^+$ and $\bfx^+$ to $\bfx^-$, respectively. Then, there exists $^{\tau''}\bfF$ on $\Omega_{\tau''}$ satisfying \eqref{eqn:pfaff_large} and
\[
^{\tau''} \bfF(\bfx^-) = ^{\tau''} \bfF(\bfx^-)\, {\sf{P}}^{\bfl_{(\bfx^-,\bfx^+)}}\,{\sf{P}}^{\bfq_{(\bfx^+,\bfx^-)}} \Longrightarrow \left[ {\sf{P}}^{\bfq_{(\bfx^+,\bfx^-)}} \right]^{-1} = {\sf{P}}^{\bfl_{(\bfx^-,\bfx^+)}},
\]
and assuming the natural definition \cite[Definition 1.4]{dol_fried} that $\left[ {\sf{P}}^{\bfq_{(\bfx^-,\bfx^+)}} \right]^{-1} = {\sf{P}}^{\bfq_{(\bfx^+,\bfx^-)}}$ we have
\[
{\sf{P}}^{\bfl_{(\bfx^-,\bfx^+)}} = {\sf{P}}^{\bfq_{(\bfx^-,\bfx^+)}}.
\]
Passing to the limit $\bfx^\pm \rightarrow \bfx$, we have that 
\begin{equation}\label{eqn:Pqx=Plx}
{\sf{P}}^{\bfq_\bfx} = {\sf{P}}^{\bfl_\bfx},
\end{equation}
where $\bfq_\bfx, \bfl_\bfx$ are parametrized, non-contractible loops in $\Omega$ starting and ending at $\bfx$ with `identical orientation' (defined by invoking the cut-surface $\tau''$ in $\Omega$ passing through $\bfx$, and both $\bfl_{\bfx}$ and $\bfq_{\bfx}$ traversing from the `$-$ side' to the `$+$ side' of $\tau''$), and both contained in $\Omega_{\tau''} \cup \{\bfx\}$.

\begin{figure}
\centering
\includegraphics[width=6.5in, height=3.0in]{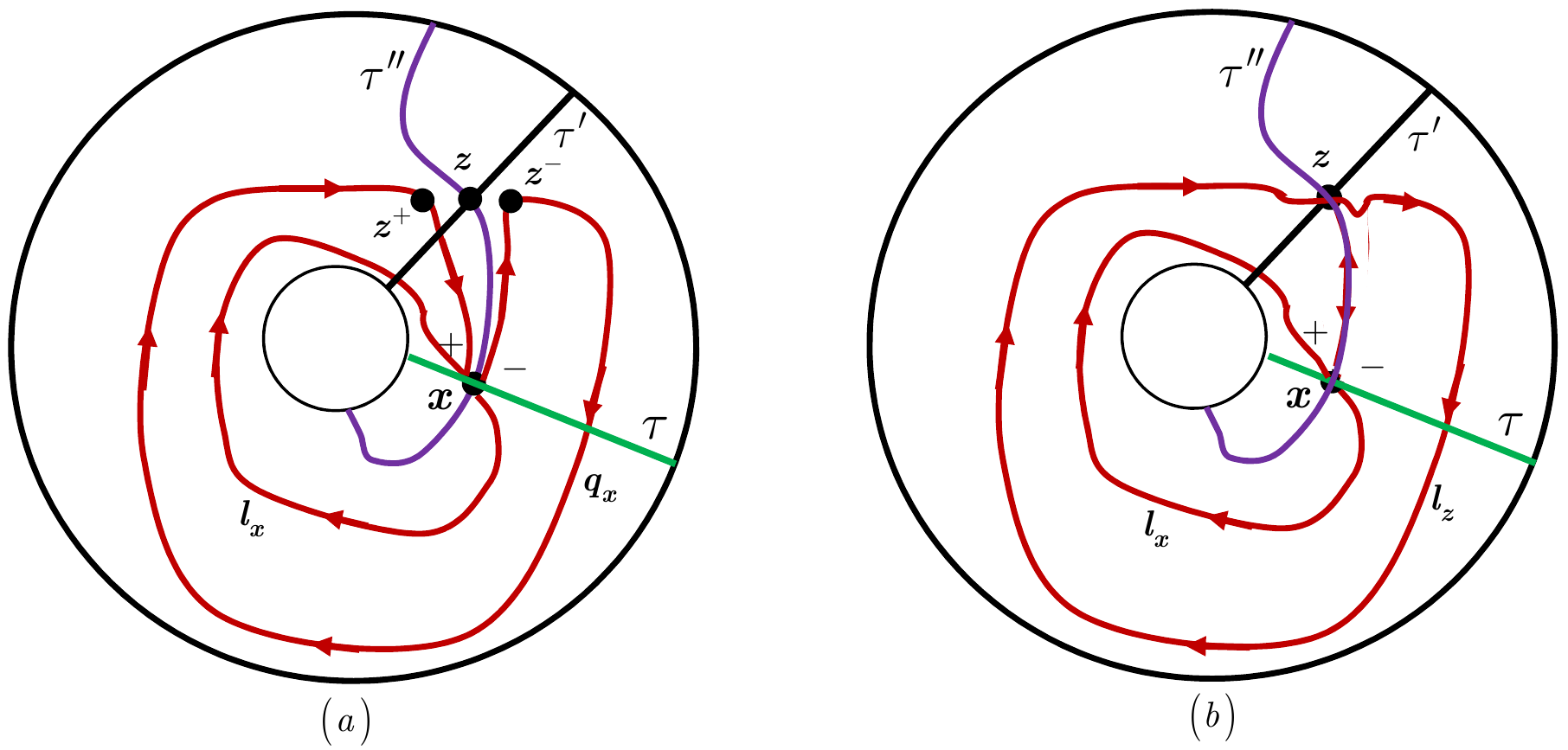}
\caption{An example when ${\sf{P}}^{\bfl_\bfx} = {\sf{P}}^{\bfq_{(\bfx,\bfz)}}\, {\sf{P}}^{\bfl_{\bfz}} \,{\sf{P}}^{\bfq_{(\bfz,\bfx)}}$ for $\bfx \in \tau$ and $\bfz \in \tau'$. ${\sf{P}}^{\bfl_\bfx} = \bfI$ iff ${\sf{P}}^{\bfl_\bfz} = \bfI$.}
\label{fig:P=I}
\end{figure}

Suppose now we introduce points $\bfz^+$ and $\bfz^-$ on the curve $\bfq_{(\bfx^-, \bfx^+)}$ as shown in Figure \ref{fig:P=I}a. The product integral has the multiplicative property
\[
{\sf{P}}^{\bfq_{(\bfx^-,\bfx^+)}} = {\sf{P}}^{\bfq_{(\bfx^-,\bfz^-)}}\,{\sf{P}}^{\bfq_{(\bfz^-,\bfz^+)}}{\sf{P}}^{\bfq_{(\bfz^+,\bfx^+)}}
\]
(see \cite[Theorem 1.5, p.11]{dol_fried}). Now choose a sequence of curves based on $\bfq_{(\bfx^-, \bfx^+)}$  in such a way that the segments $\bfq_{(\bfx^-,\bfz^-)}$ and $\bfq_{(\bfz^+,\bfx^+)}$ approach the curve joining $\bfx$ to $\bfz$, the latter entirely contained in $\tau''$, but with opposite sense of traversal and in the limit we have
\begin{equation}\label{eqn:Pqx=Pqz}
{\sf{P}}^{\bfq_\bfx} = {\sf{P}}^{\bfq_{(\bfx,\bfz)}}\, {\sf{P}}^{\bfl_{\bfz}} \,{\sf{P}}^{\bfq_{(\bfz,\bfx)}},
\end{equation}
where $\bfl_\bfz$ is a non-contractible loop in $\Omega$ starting and ending at $\bfz$, running from the `$-$' side to the `$+$' side of $\tau''$, and contained in $\Omega_{\tau''} \cup \{\bfz\}$, and we note that $\left[ {\sf{P}}^{\bfq_{(\bfx,\bfz)}} \right]^{-1} = {\sf{P}}^{\bfq_{(\bfz,\bfx)}}$.

It seems natural (and we assume this) that $\bfl_\bfx$ can be chosen in such a way that it intersects both $\tau$ and $\tau''$ exactly once at $\bfx$ and, similarly, $\bfl_\bfz$ can be chosen in such a way that it intersects both $\tau'$ and $\tau''$ exactly once at $\bfz$ (see Fig. \ref{fig:P=I}b). Then $^\tau \bfQ = \bfI$ implies that ${\sf{P}}^{\bfl_\bfx} = \bfI$ (since ${^{\tau}\bfF} (\bfx^+) = {^{\tau}\bfF} (\bfx^-)\, {\sf{P}}^{\bfl_{(\bfx^-,\bfx^+)}}$), which, along with \eqref{eqn:Pqx=Plx}, implies ${\sf{P}}^{\bfq_\bfx} = \bfI$, and then it can be inferred, using \eqref{eqn:Pqx=Pqz}, that $^{\tau'} \bfQ = \bfI$ (since ${^{\tau'}\bfF} (\bfz^+) = {^{\tau'}\bfF} (\bfz^-)\, {\sf{P}}^{\bfq_{(\bfz^-,\bfz^+)}}$).
\begin{remark}
For any solution $\left({^\tau\bfy}, {^\tau\bfF}\right)$ of \eqref{eqn:pfaff_large} on $\Omega_\tau$, $^{\tau} \bfQ(\bfx) = \bfI$ if and only if ${\sf{P}}^{\bfl_\bfx} = \bfI$, for any $\bfx \in \tau$ a cut-surface of $\Omega$ and $\bfl_\bfx$ a parametrized, non-contractible loop in $\Omega$ starting and ending at $\bfx$. 

For $\bfl_\bfx$ running from the `$-$' side to the `$+$' of $\tau$,
\begin{equation}\label{rem:4.7}
\begin{split}
& ^{\tau} \bfQ = {^\tau \bfF^-}(\bfx) {\sf{P}}^{\bfl_\bfx} \left[ {^\tau \bfF^-}(\bfx) \right]^{-1},\\
& \left[ ^\tau \bfR^- (\bfx) \right]^T {^\tau \bfQ} \left[ ^\tau \bfR^-(\bfx) \right] = \bfU(\bfx){\sf{P}}^{\bfl_\bfx}\left[ \bfU(\bfx)\right]^{-1}
\end{split}
\end{equation}
for all $\bfx \in \tau$. The function on the right hand side of \eqref{rem:4.7}$_1$ is constant on $\tau$, even though ${\sf{P}}^{\bfl_\bfx}$, as $\bfx$ varies on $\tau$, is not, and neither is $\bfU(\bfx){\sf{P}}^{\bfl_\bfx}\left[ \bfU(\bfx)\right]^{-1}$, where $\bfU$ is the right stretch tensor field of $\bfC = \bfU^2$ on $\Omega$. The work of Shield \cite{shield}, along with the use of the product integral, can provide an explicit characterization of the rotation field, $^\tau \bfR$, of $^\tau \bfy$ on $\Omega_\tau$ (and hence the values $^\tau \bfR^-$) in terms of the prescribed strain field and an `initial condition'; \eqref{rem:4.7}$_2$ also provides a characterization of the variation of the $^\tau\bfR^-$ field on $\tau$ in terms of given data, with an explicit indication of the nonlocality involved.
\end{remark}

\section*{Acknowledgments}
I am very grateful to Reza Pakzad and Raz Kupferman for their valuable comments and discussion. I thank Reza for reading the whole paper and Raz for taking a look at the finite deformation part. I also thank Reza for showing me alternative proofs of the main results of this paper for bodies in two space dimensions with a single hole without involving any line or product integrals, but capitalizing only on appropriate statements of rigidity. I acknowledge the support of the Center for Nonlinear Analysis at Carnegie Mellon and grants ARO W911NF-15-1-0239 and NSF-CMMI-1435624.

\section*{Appendix}

\appendix

\section{Strain Compatibility on a simply connected region}\label{comp_sc}
 For the sake of completeness we collect some classical results related to questions of strain compatibility in the following sections. \emph{All regions considered in these appendices are simply connected, unless mentioned otherwise}. We repeatedly use the argument that if $f_{,k} = 0$ on the domain then $f$ is a constant, which uses the fact that the region in question is path connected.

\subsection{Small deformation}\label{comp_small}
\begin{theorem}
Given a $C^2$ second-order tensor field $\bfveps$ on a simply connected region, it is necessary and sufficient for the existence of a $C^3$ displacement field $\bfu$ on it satisfying $(grad \,\bfu)_{sym} = \bfveps$ that the St.-Venant compatibility condition \eqref{eqn:loc_comp}$_3$ be satisfied.
\end{theorem}

\emph{Proof}: \emph{Necessity} - $u_i$ exists satisfying
\begin{equation}\label{eqn:a1_0}
\begin{split}
& u_{i,k} = \veps_{ik} + \omega_{ik};\ \ \ \veps_{ik} = \frac{1}{2} \left( u_{i,k} + u_{k,i} \right); \ \ \ \omega_{ik} := \frac{1}{2} \left( u_{i,k} - u_{k,i} \right). \\
& \omega_{ik,l} = \frac{1}{2} \left( u_{i,kl} + u_{l,ki} - u_{l,ki} - u_{k,li}\right) = e_{il,k} - e_{kl,i}.
\end{split}
\end{equation}
The infinitesimal rotation $\bfomega$ is twice continuously differentiable and therefore, $\omega_{ik,lm} - \omega_{ik,ml} = 0$ implies \eqref{eqn:loc_comp}$_3$.

\emph{Sufficiency} - We assume \eqref{eqn:loc_comp}$_3$ is satisfied and define
\begin{equation}\label{eqn:a1_1}
\begin{split}
& E_{ikl} := \veps_{il,k} - \veps_{kl,i},\\
& \omega_{ik} (\bfx) = W_{ik} + \int_{\bfc{(\bfx_0,\bfx)}} E_{ikl} (\bfc)\, dc_l,
\end{split}
\end{equation}
where $\bfc_{(\bfx_0,\bfx)}$ is some path in the region from $\bfx_0$ to $\bfx$ and $\bfW$ is a arbitrary skew-symmetric tensor. Since $E_{ikl,m} = E_{ikm,l}$ due to \eqref{eqn:loc_comp}$_3$, $\bfomega$ as defined in \eqref{eqn:a1_1}$_2$ is independent of path, and defines a smooth function that satisfies 
\begin{equation}\label{eqn:a1_2}
\omega_{ik,l} = E_{ikl}
\end{equation}
on the region which is unique up to the choice of $\bfomega(\bfx_0) = \bfW$. Since \eqref{eqn:a1_2} and \eqref{eqn:a1_1}$_1$ imply
\[
\veps_{ik,l} + \omega_{ik,l} - \left( \veps_{il,k} + \omega_{il,k} \right) = 0,
\]
the definition
\[
u_i(\bfx) = u^0_i + \int_{\bfc_{(\bfx_0,\bfx)}} \left( \veps_{ik} (\bfc) + \omega_{ik}(\bfc) \right)\, dc_k,
\]
for any $\bfu^0$ an arbitrary constant vector, is independent of the path $\bfc_{(\bfx_0,\bfx)}$ chosen to connect $\bfx_0$ and $\bfx$, and hence defines a smooth displacement field whose symmetrized gradient equals the given $\bfveps$ field.
\subsection{Rigidity for smooth infinitesimal deformations}\label{rigid_small}
\begin{theorem}
If two $C^1$ displacement fields on a region have identical strain fields, then they differ at most by an infinitesimally rigid deformation. The region here need not be simply-connected.
\end{theorem}

\emph{Proof}: Let $\bfu^1$ and $\bfu^2$ be the two displacement fields with strain fields $\bfveps^1$ and $\bfveps^2$ and rotation field $\bfomega^1$ and $\bfomega^2$, respectively, defined from the corresponding displacement fields by relations \eqref{eqn:a1_0} and let $\bfveps^1 = \bfveps^2$. Then
\begin{equation*}
 \omega^1_{ij,k} - \omega^2_{ij,k} = \veps^1_{ik,j} - \veps^2_{jk,i} = 0,
\]
using a similar computation as in \eqref{eqn:a1_0}, and therefore $\bfW : = \bfomega^1 - \bfomega^2$ is a constant skew-symmetric tensor on the region (which is path-connected). We then have
\[
u^1_{i,j} - u^2_{i,j} = W_{ij},
\]
and integrating along paths from an arbitrarily fixed $\bfz$ to all points $\bfx$ in the region and rearranging terms we obtain
\[
\bfu^1(\bfx) - \bfu^1(\bfz) = \bfu^2(\bfx) - \bfu^2(\bfz) + \bfW \left[ \bfx -\bfz \right].
\]
Since $\bfz$ was arbitrarily fixed, this implies that $\bfu^1$ and $\bfu^2$ are related by an infinitesimally rigid deformation by Definition \ref{def:rigid_small}.
\subsection{Finite deformation}\label{comp_large}
The treatment here is from Sokolnikoff \cite{sokol}. While the considerations below relate to fundamental relations in Riemannian geometry, we intentionally emphasize the purely algebraic fact that if the functions $h,g,x,y$ (defined below)  satisfy \eqref{eqn:metric_rel}, then they necessarily satisfy \eqref{eqn:christoff_rel} and the definition \eqref{eqn:christoff_1} implies \eqref{eqn:metric_cov_const}.

Let $\Omega_x$ and $\Omega_y$ be two $3$-d coordinate patches, i.e., open bounded regions of $\R^3$, and let $y:\Omega_x \rightarrow \Omega_y$ be a $C^2$ diffeomorphism with a $C^2$ inverse that we denote by $x:\Omega_y \rightarrow \Omega_x$. Let $h:\Omega_y \rightarrow \R^{3\times3}$ and $g:\Omega_x \rightarrow \R^{3\times3}$ be two prescribed matrix fields with range in the set of symmetric, positive definite $3\times3$ matrices. In the following all indices range over the set $\{ 1,2,3\}$. Assume
\begin{equation}\label{eqn:metric_rel}
h_{ij} = x^\alpha_{,i} \left( g_{\alpha\beta} \circ x \right) x^\beta_{,j} \Longleftrightarrow g_{\alpha \beta} =  y^i_{,\alpha} \left( h_{ij} \circ y \right)  y^j_{,\beta}
\end{equation}
holds. Then, using the symmetry of $g_{\alpha \beta}$ (and switching some dummy indices),
\begin{equation*}
\begin{split}
 h_{ij,k} & = \left( x^\alpha_{,ik} x^\beta_{,j} + x^\beta_{,i} x^\alpha_{,jk} \right) \left( g_{\alpha \beta} \circ x\right)  + x^\alpha_{,i} x^\beta_{,j} x^\gamma_{,k} \left( g_{\alpha \beta,\gamma} \circ x \right),\\
h_{ik,j} & = \left( x^\alpha_{,ij} x^\beta_{,k} + x^\beta_{,i} x^\alpha_{,kj} \right) \left( g_{\alpha \beta} \circ x\right)  + x^\alpha_{,i} x^\gamma_{,k} x^\beta_{,j} \left( g_{\alpha \gamma,\beta} \circ x \right),\\
 h_{jk,i} & = \left( x^\alpha_{,ji} x^\beta_{,k} + x^\beta_{,j} x^\alpha_{,ki} \right) \left( g_{\alpha \beta} \circ x\right)  + x^\beta_{,j} x^\gamma_{,k} x^\alpha_{,i} \left( g_{\beta \gamma,\alpha} \circ x \right).
\end{split}
\]
Now define $\hat{\Gamma}:\Omega_x \rightarrow \R^{3 \times 3 \times 3}$ and $\hat{H}:\Omega_y \rightarrow \R^{3 \times 3 \times 3}$ as
\begin{equation}\label{eqn:christoff_1}
\begin{split}
\hat{H}_{ijk}&  := \half \left( h_{ik,j} + h_{jk,i} - h_{ij,k} \right),\\
\hat{\Gamma}_{\alpha \beta \gamma} & := \half \left( g_{\alpha \gamma,\beta} + g_{\beta \gamma,\alpha} - g_{\alpha \beta,\gamma} \right).
\end{split}
\end{equation}
Then,
\[
\hat{H}_{ijk} = x^\alpha_{,i} x^\beta_{,j} x^\gamma_{,k} \left( \hat{\Gamma}_{\alpha \beta \gamma} \circ x \right) + x^\alpha_{,ji} x^\beta_{,k} \left( g_{\alpha \beta} \circ x \right). 
\]
Next we define $H: \Omega_y \rightarrow \R^{3 \times 3 \times 3}$ and $\Gamma: \Omega_x \rightarrow \R^{3 \times 3 \times 3}$ by 
\[
H^k_{ij} = h^{km} \hat{H}_{ijm} \ \ \mbox{and} \ \  \Gamma^\rho_{\alpha \beta} = g^{\rho \nu} \hat{\Gamma}_{\alpha \beta \nu},
\]
where $h^{km}$ and $g^{\alpha \beta}$ are the components of the matrices $h^{-1}$ and $g^{-1}$, respectively. We note that $\hat{H}, H, \hat{\Gamma}, \Gamma$ are all symmetric in their lower first two indices. Noting from \eqref{eqn:metric_rel} that
\[
h^{km} = \left( g^{\alpha \beta} \circ x \right) \left( y^k_{,\alpha} \circ x \right) \left( y^m_{,\beta} \circ x \right),
\]
we obtain
\[
H^k_{ij} = \left( y^k_{,\alpha} \circ x \right) x^\alpha_{,ji} + \left( y^k_{,\rho} \circ x \right) x^\alpha_{,i} x^\beta_{,j} \left( \Gamma^\rho_{\alpha \beta} \circ x \right),
\]
which, after rearrangement of terms, yields
\begin{equation}\label{eqn:a3_misc}
x^\mu_{,ji} = x^\mu_{,k} H^k_{ij} - x^\alpha_{,i} x^\beta_{,j} \left( \Gamma^\mu_{\alpha \beta} \circ x \right).
\end{equation}
Of course, the computations above indicate that interchanging the list $\left( x,y, g,g^{-1}, h, h^{-1}, \hat{\Gamma}, \hat{H}, \Gamma, H \right)$ by $\left( y,x, h, h^{-1}, g,g^{-1}, \hat{H}, \hat{\Gamma}, H, \Gamma \right)$ in the above formulae is admissible. We thus have
\begin{equation}\label{eqn:christoff_rel}
y^i_{,\alpha \beta} = y^i_{,\gamma} \Gamma^\gamma_{\alpha \beta} - y^j_{,\alpha} y^k_{,\beta} \left( H^i_{jk} \circ y \right).
\end{equation}

Another result we will need for our compatibility argument to follow is as follows:
\begin{equation*}
\begin{split}
\hat{\Gamma}_{\alpha \gamma \beta} + \hat{\Gamma}_{\beta \gamma \alpha} & = g_{\alpha \beta,\gamma},\\
\left(g_{\alpha \beta} g^{\beta \mu} \right)_{,\gamma} & = 0 
\end{split}
\end{equation*}
so that
\[
g^{\rho \mu}_{,\gamma} = - g^{\rho \alpha} \left( \hat{\Gamma}_{\alpha \gamma \beta} + \hat{\Gamma}_{\beta \gamma \alpha} \right) g^{\beta \mu} = - \left( g^{\rho \alpha} \Gamma^\mu_{\alpha \gamma} + g^{\beta \mu} \Gamma^\rho_{\beta \gamma} \right),
\]
which implies
\begin{equation}\label{eqn:metric_cov_const}
g^{\alpha \beta}_{,\mu} = - \left( g^{\alpha \gamma} \Gamma^\beta_{\gamma \mu} + g^{\gamma \beta} \Gamma^\alpha_{\gamma \mu} \right).
\end{equation}

Let $\Omega$ be a simply connected region and let $\bfC: \Omega \rightarrow {\cal{P}}_{sym}$ be a prescribed $C^2$ field on $\Omega$, where ${\cal{P}}_{sym}$ is the set of all positive-definite, symmetric tensors on $V_3$. Let $\Omega_x$ be a Rectangular Cartesian coordinate patch parametrizing $\Omega$ and $C:\Omega \rightarrow \R^{3 \times 3}$ be the component map of $\bfC$ with respect to the rectangular Cartesian basis of the parametrization of $\Omega$ by $\Omega_x$.
\begin{theorem}
If there exists a deformation $\bfy: \Omega \rightarrow {\cal{E}}_3$ satisfying 
\begin{equation}\label{eqn:a3_3}
\left( grad \, \bfy \right)^T grad \bfy = \bfC,
\end{equation}
then \eqref{eqn:loc_comp}$_2$ is satisfied in $\Omega$. Conversely, if the matrix field $C$ satisfies \eqref{eqn:loc_comp}$_2$, then there exists a deformation $\bfy: \Omega \rightarrow {\cal{E}}_3$ that satisfies \eqref{eqn:a3_3}.
\end{theorem}
\emph{Proof}: \emph{Necessity} of \eqref{eqn:loc_comp}$_2$ for \eqref{eqn:a3_3} - Let $y: \Omega_x \rightarrow \Omega_y \subset \R^3$ be the coordinate map representing the parametrization of $\bfy(\Omega)$ by the same Rectangular Cartesian system for ${\cal{E}}_3$ used to parametrize $\Omega$. Then
\[
 y^i_{,\alpha} \, \delta_{ij} \, y^j_{,\beta} = C_{\alpha \beta}
\]
holds. Making the identification of $h = I$ and $g = C$ in \eqref{eqn:metric_rel}, we have $H = 0$ and \eqref{eqn:christoff_rel} implies
\begin{equation}\label{eqn:a3_4}
\begin{split}
y^i_{,\alpha} & = F^i_\alpha,\\
F^i_{\alpha,\beta} & = \Gamma^\gamma_{\alpha \beta} \, F^i_\gamma.
\end{split}
\end{equation}
Due to the smoothness of $\bfC$ and \eqref{eqn:a3_4} we obtain
\begin{equation*}
\begin{split}
& y^i_{\alpha,\beta \rho} = y^i_{\alpha, \rho \beta}\\
\Longrightarrow \, &  F^i_{\gamma,\rho} \, \Gamma^\gamma_{\alpha \beta} + F^i_\gamma \, \Gamma^\gamma_{\alpha \beta,\rho} - F^i_{\gamma,\beta} \, \Gamma^\gamma_{\alpha \rho} - F^i_\gamma \, \Gamma^\gamma_{\alpha \rho, \beta} = 0\\
\Longrightarrow \, & F^i_\mu \left( \Gamma^\mu_{\alpha \beta,\rho} - \Gamma^\mu_{\alpha \rho,\beta} + \Gamma^\mu_{\gamma \rho} \Gamma^\gamma_{\alpha \beta} - \Gamma^\mu_{\gamma \beta} \Gamma^\gamma_{\alpha \rho} \right) = 0
\end{split}
\end{equation*}
and since $F$ is an invertible matrix (due to $\bfC \in {\cal{P}}_{sym}$), \eqref{eqn:loc_comp}$_2$ holds.

\emph{Sufficiency} of \eqref{eqn:loc_comp}$_2$ for \eqref{eqn:a3_3} - By a theorem of Thomas \cite{thomas} (also the Froebenius theorem in the differential geometry literature), we have that a solution to \eqref{eqn:a3_4} exists, with freely specifiable value of $F$ and $y$ at arbitrarily chosen points of $\Omega_x$, if $\Gamma^\gamma_{\alpha \beta} = \Gamma^\gamma_{\beta \alpha}$ (which holds by definition of $\Gamma$), and \eqref{eqn:loc_comp}$_2$ hold. 

Specify the value of $F$ at an arbitrarily chosen point $x_0 \in \Omega_x$ such that $F^i_\alpha (x_0) F^i_\beta (x_0) = C_{\alpha \beta} (x_0)$, written alternatively as $F_0^T F_0 = C_0 \Longrightarrow F_0 C^{-1}_0 F_0^T = I$.  For an  $F$ field satisfying \eqref{eqn:a3_4} with the identification $C = g$ in \eqref{eqn:christoff_1} and \eqref{eqn:metric_cov_const}, consider now
\begin{equation*}
\begin{split}
\left( C^{\alpha \beta} F^i_\alpha F^j_\beta \right)_{,\mu}  = \left( C^{\alpha \beta}_{,\mu} + C^{\gamma \beta} \Gamma^\alpha_{\gamma \mu} + C^{\alpha \gamma} \Gamma^\beta_{\gamma \mu} \right) F^i_\alpha F^j_\beta = 0,
\end{split}
\end{equation*}
by \eqref{eqn:metric_cov_const}. Thus we have that
\[
FC^{-1}F^T = I \Longrightarrow C = F^TF \ \ \ \mbox{on}\ \ \Omega_x,
\]
and defining $\bfy = y^i \bfe_i$ where $\bfe_i, i = 1,2,3$ is the (spatially constant) natural basis of the Rectangular Cartesian coordinate system used to parametrize $\Omega$, we find that $\bfy$ satisfies \eqref{eqn:a3_3}.
\subsection{Rigidity for smooth finite deformations}\label{rigid_large}
Let $\bfy^*:\Omega \rightarrow {\cal{E}}_3$ and $\bfz:\Omega \rightarrow {\cal{E}}_3$ be two $C^2$ deformations of $\Omega$. Consider a rectangular Cartesian parametrization of ${\cal{E}}_3$ under which $\Omega$ maps to the set of coordinates $\Omega_x \subset \R^3$, $\bfy^*(\Omega)$ maps to $\Omega_y$, and $\bfz(\Omega)$  maps to $\Omega_z$. Let $y^*:\Omega_x \rightarrow \Omega_y$, $z:\Omega_x \rightarrow \Omega_z$, $y:\Omega_z \rightarrow \Omega_y$, and $x: \Omega_z \rightarrow \Omega_x$ be the corresponding deformations, represented in coordinates. $\Omega$ need not be simply connected.

\begin{theorem}
If $\bfy^*$ and $\bfz$ have the same right Cauchy-Green tensor fields then they are rigidly related to each other.
\end{theorem}

\emph{Proof}: We have, following \cite{shield},
\begin{equation}\label{eqn:a4_1}
\begin{split}
& y^{*i}_{,j} y^{*i}_{,k} = z^m_{,j}z^m_{,k} = C_{jk}\\
\Rightarrow & \left( y^i_{,m} \circ z \right) z^m_{,j} \left( y^i_{,n} \circ z \right) z^n_{,k} = z^m_{,j} z^m_{,k}\\
\Rightarrow \, & y^i_{,p} y^i_{,q} =  y^\alpha_{,p} \, \delta_{\alpha \beta} \, y^\beta_{,q} = x^j_{,p} \left( z^m_{,j} \circ x \right) \left( z^m_{,k} \circ x \right) x^k_{,q} = \delta^m_p \delta^m_q = \delta_{pq},\\
\end{split}
\end{equation}
and identifying $x$, $h$, and $g \circ x$ in \eqref{eqn:metric_rel} with $y$, $I$, and $I$, respectively, here, we have from \eqref{eqn:a3_misc} that
\[
y^\alpha_{,pq} = 0 \ \ \mbox{on} \ \ \Omega_z.
\]
Due to the path connectedness of $\Omega_z$ (induced from $\Omega_x$), this implies that there exists a constant matrix $R$ satisfying $y^\alpha_{,p} = R^\alpha_p$ on $\Omega_z$ with $R^TR = I$ from \eqref{eqn:a4_1}$_3$. Integrating along an arbitrarily chosen path from $z(w_0) \in \Omega_z$ to $z(w) \in \Omega_z$ for $w_0, w \in \Omega_x$, we obtain
\[
y^\alpha \circ z (w) = y^\alpha \circ z (w_0) + R^\alpha_p \left[ z^p(w) - z^p(w_0) \right] \Longrightarrow y^*(w) = y^*(w_0) + R \left[ z(w) - z(w_0) \right]
\]
for any $w \in \Omega_x$ and $w_0 \in \Omega_x$ and therefore $\bfy^*$ and $\bfz$ are rigidly related to each other by Definition \ref{def:rigid_large}.

\bibliographystyle{alpha}\bibliography{weingarten}

\end{document}